\documentclass[fleqn,usenatbib]{mnras}

\usepackage{amssymb}	% Extra maths symbols

\usepackage{newtxtext,newtxmath}
\usepackage{indentfirst}
\usepackage[T1]{fontenc}

\DeclareRobustCommand{\VAN}[3]{#2}
\let\VANthebibliography\thebibliography
\def\thebibliography{\DeclareRobustCommand{\VAN}[3]{##3}\VANthebibliography}

\usepackage{graphicx}	% Including figure files
\usepackage{amsmath}	% Advanced maths commands
\usepackage{bm}
\usepackage{hyperref}

\newcommand{\aref}[1]{\hyperref[#1]{Appendix~\ref{#1}}}

\usepackage{scalerel,tikz}
\usetikzlibrary{svg.path}
\definecolor{orcidlogocol}{HTML}{A6CE39}
\tikzset{orcidlogo/.pic={
 \fill[orcidlogocol] svg{M256,128c0,70.7-57.3,128-128,128C57.3,256,0,198.7,0,128C0,57.3,57.3,0,128,0C198.7,0,256,57.3,256,128z};
 \fill[white] svg{M86.3,186.2H70.9V79.1h15.4v48.4V186.2z}
 svg{M108.9,79.1h41.6c39.6,0,57,28.3,57,53.6c0,27.5-21.5,53.6-56.8,53.6h-41.8V79.1z M124.3,172.4h24.5c34.9,0,42.9-26.5,42.9-39.7c0-21.5-13.7-39.7-43.7-39.7h-23.7V172.4z}
 svg{M88.7,56.8c0,5.5-4.5,10.1-10.1,10.1c-5.6,0-10.1-4.6-10.1-10.1c0-5.6,4.5-10.1,10.1-10.1C84.2,46.7,88.7,51.3,88.7,56.8z};
}}
\newcommand\orcidicon[1]{\href{https://orcid.org/#1}{\mbox{\scalerel*{
\begin{tikzpicture}[yscale=-1,transform shape]
\pic{orcidlogo};
\end{tikzpicture}
}{|}}}}

\title{Detection of metallicity correlations in 100 nearby galaxies}

\author[Li et al.]{Zefeng Li$^{\orcidicon{0000-0001-7373-3115}}$, $^1$\thanks{E-mail: zefeng.li@anu.edu.au}
Mark R. Krumholz$^{\orcidicon{0000-0003-3893-854X}}$, $^{1, 2}$
Emily Wisnioski$^{\orcidicon{0000-0003-1657-7878}}$, $^{1, 2}$
J. Trevor Mendel$^{\orcidicon{0000-0002-6327-9147}}$, $^{1, 2}$
\newauthor Lisa J. Kewley$^{\orcidicon{0000-0001-8152-3943}}$, $^{1, 2}$
Sebastian F. S\'anchez$^{\orcidicon{0000-0001-6444-9307}\, 3}$
and Llu\'is Galbany$^{\orcidicon{0000-0002-1296-6887}\, 4}$\\
$^1$Research School of Astronomy \& Astrophysics, Australian National University, Canberra, 2612 ACT, Australia\\
$^2$ARC Centre of Excellence for All Sky Astrophysics in 3 Dimensions (ASTRO 3D)\\
$^3$Instituto de Astronom\'ia, Universidad Nacional Auton\'oma de M\'exico, A.P. 70-264, 04510 M\'exico, D.F., Mexico.\\
$^4$Departamento de F\'isica Te\'orica y del Cosmos, Universidad de Granada, E-18071 Granada, Spain
}

\date{Accepted 2021 April 28. Received 2021 April 8; in original form 2021 January 28}

\pubyear{2021}

\begin{document}
\label{firstpage}
\pagerange{\pageref{firstpage}--\pageref{lastpage}}
\maketitle

\begin{abstract}
In this paper we analyse the statistics of the 2D gas-phase oxygen abundance distributions of 100 nearby galaxies drawn from the Calar Alto Legacy Integral Field spectroscopy Area survey. After removing the large-scale radial metallicity gradient, we compute the two-point correlation functions of the resulting metallicity fluctuation maps. We detect correlations in the majority of our targets, which we show are significantly in excess of what is expected due to beam-smearing, and are robust against the choice of metallicity diagnostic. We show that the correlation functions are generally well-fit by the predictions of a simple model for stochastic metal injection coupled with diffusion, and from the model we show that, after accounting for the effects of both beam smearing and noise, the galaxies in our sample have characteristic correlation lengths of $\sim1$ kpc. Correlation lengths increase with both stellar mass and star formation rate, but show no significant variation with Hubble type, barredness, or merging state. We also find no evidence for a theoretically-predicted relationship between gas velocity dispersion and correlation length, though this may be due to the small dynamic range in gas velocity dispersion across our sample. Our results suggest that measurements of 2D metallicity correlation functions can be a powerful tool for studying galaxy evolution.
\end{abstract}


%%%%%%%%%%%%%%%%%%%%%%%%%%%%%%%%%%%%%%%%%%%%%%%%%%

%%%%%%%%%%%%%%%%% BODY OF PAPER %%%%%%%%%%%%%%%%%%

\section{Introduction}

Gas-phase oxygen abundance (metallicity) is a tracer of galactic and chemical evolution in Galactic and extra-galactic astronomy \citep[for reviews, see][]{K19a, Maiolino19, Sanchez20a}. Traditionally single-fibre spectroscopic observations have been used to explore the global or central metallicities of galaxies \citep[SDSS; e.g.][]{Tremonti04}, while long-slit spectra have been used to study metallicity gradients \citep[e.g.][]{Henry99}. Thanks to integral field spectroscopy (IFS), however, the era has come when we can make resolved spectroscopic observations of H~\textsc{ii} regions in galaxies beyond the Milky Way, rather than measuring a single average metallicity, or, at best, measuring a radial gradient. The gas-phase metallicity of galaxy discs can now be resolved for large numbers of galaxies, using the current generation IFS surveys such as the Calar Alto Legacy Integral Field spectroscopy Area survey \citep[CALIFA;][]{Sanchez12, Sanchez14}, the Sydney-Australian-Astronomical-Observatory Multi-object Integral-Field Spectrograph survey \citep[SAMI;][]{Croom12}, the Mapping Nearby Galaxies at APO survey \citep[MaNGA;][]{Bundy15}, and the MUSE surveys (e.g. MUSE Atlas Discs, MAD: \citealt{Erroz-Ferrer19}; PHANGS-MUSE: E. Schinnerer, 1100.B-0651; AMUSING++: \citealt{Lopez20}). Based on these surveys a number of authors have published full 2D metallicity maps for nearby galaxies \citep[e.g.][]{Rosales-Ortega11, Sanchez-Menguiano16_conf, Sanchez-Menguiano16, Sanchez-Menguiano17, Sanchez-Menguiano20a, Sanchez-Menguiano20b}. Azimuthal averages of these maps typically yield negative (ranging from 0.0 to $\sim-0.1$ dex/kpc) metallicity gradients \citep[e.g.][]{Zinchenko16, Belfiore17, Ho18, Poetrodjojo18, Sanchez-Menguiano18, Kreckel19}. However, there has been more limited exploration of the statistics of the full 2D, non-averaged maps. One exception is the recent study by \citet{Kreckel20}, who carry out a quantitative analysis of a full 2D metallicity distribution a sample of galaxies drawn from the PHANGS sample, observed with MUSE. However, their sample is limited to eight galaxies.

These studies lead to a next-generation question: what else can a 2D metallicity map tell about the processes by which heavy elements are produced in and transported through galaxies, and how these processes correlate with other galactic properties, e.g., stellar mass ($M_*$) or star formation rate (SFR)? Part of the reason why this question has remained unanswered is that most existing theoretical models are unable to predict 2D metal distributions. In principle such information could be extracted from cosmological simulations. However, in practice such studies are limited by resolution. For example, \citet{Kobayashi11} carry out chemodynamic simulations of a Milky-Way-type galaxy, including supernova feedback and chemical enrichment, and predict the spatial distribution of elements from oxygen to zinc. However, their mass resolution is $>10^5$ M$_\odot$, which, at the mean density $\sim 1$ H cm$^{-3}$ of the Milky Way's ISM, corresponds to a mean inter-particle spacing $\gtrsim 150$ pc -- sufficient to separate the disc, bulge, and halo (which is the goal of their study), but not sufficient to look at distributions within the disc. Similarly, \citet{Minchev13} carry out chemodynamic simulations, but supplement these with a 1D subgrid model to follow the chemistry of the (unresolved) thin disc, eliminating the possibility of studying 2D distributions. The highest resolution cosmological chemodynamic simulations carried out to date reach mass resolutions of a $\mbox{few} \times 10^3$ M$_\odot$ for Milky Way-sized galaxies \citep[e.g.,][]{Escala18a}, but even this only gives a spatial resolution $\sim 100$ pc. 

On the other hand, higher resolution is possible in simulations that do not attempt to follow the Milky Way-sized galaxies from formation all the way to $z=0$. Simulations of this type have made important advances in understanding the basic mechanisms for metal transport, including supernova-driven turbulence \citep{deAvillez02, Colbrook17a}, thermal instability \citep{Yang12}, gravitational instability \citep{Petit15a}, and dilution by cosmological accretion \citep{Ceverino16a}. However, while these studies can provide useful estimates of the rates at which various processes transport metals, they lack the cosmological context required to predict the full metal distribution.

\citet[][hereafter KT18]{KT18} recently proposed a quantitative model that attempts to distill the basic results from the simulations. This model is based on stochastically forced diffusion, and predicts the multiscale statistics of metal fields that result from the competition between mixing and metal production. Despite its simplicity, this model establishes the principle that the statistics of the metal distribution encode a great deal of information about metal production and transport, and its predictions are at least qualitatively consistent with the distributions observed by \citet{Kreckel20}. However, making use of this tool will require both more theoretical work and observational analysis of substantially larger samples.

In this paper we begin this project by studying the metallicity distributions for a sample of 100 galaxies observed as part of the CALIFA survey, spanning a broad range of properties (e.g. stellar mass, star formation rate). We make full use of 2D metal maps and present a quantitative statistical analysis using two-point correlation function that enables us to study how diffusive processes mix metals and shape the metallicity distributions of galaxies, as a function of global galaxy properties. This work is a first step toward answering questions as to what the full 2D metal distributions tell us about the history of star formation and galaxy formation.

The outline of this paper is as follows. In \autoref{sec:data}, we give an overview of CALIFA catalogue data of which we make use, including a discussion of possibly-important observational biases. In \autoref{sec:methodology}, we discuss our method for analysing the metallicity distributions. In \autoref{sec:results} we introduce the correlation length which is the fundamental quantity that we extract from the maps, and discuss its dependency on galaxy properties. Finally we compare our results with previous work and draw conclusions in \autoref{sec:comparison} and \autoref{sec:conclusions}. Throughout this paper we use a cosmology defined by $\rm{H}_0 = 67.8\ \rm{km\ s} ^{-1} \rm{Mpc}  ^{-1}$ and $\Omega_{\Lambda} = 0.692$ \citep{Planck16}.

\section{Data}
\label{sec:data}

\begin{figure*}
\includegraphics[width=1.0\linewidth]{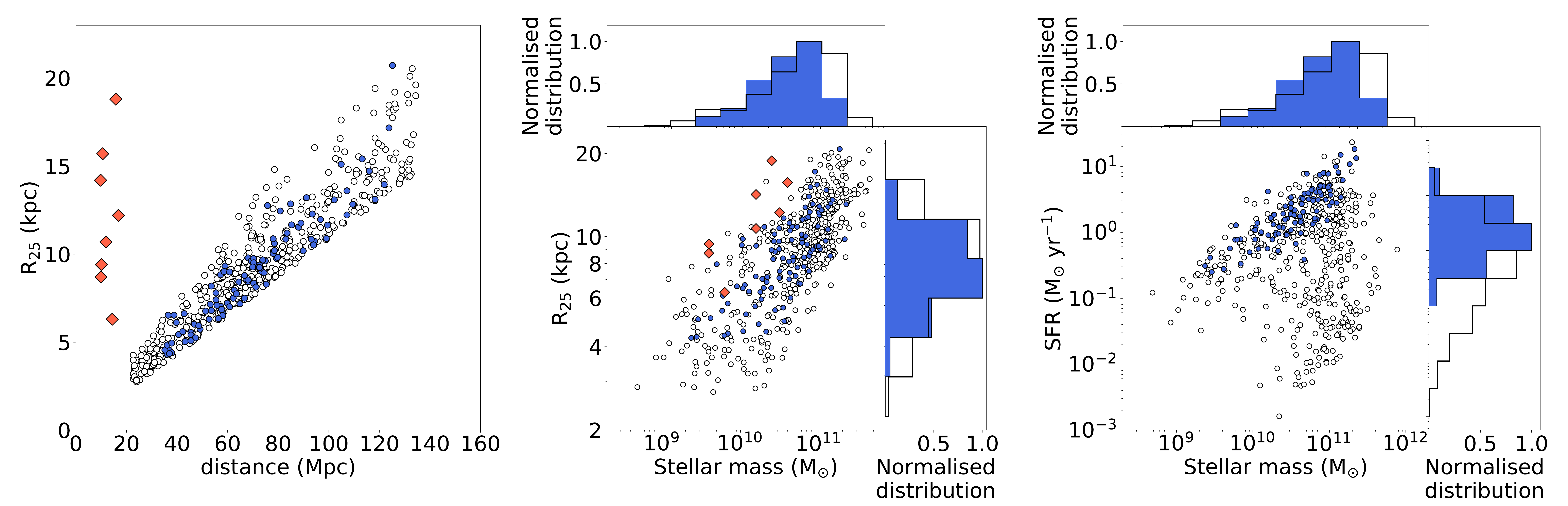}
\caption{Left: $r$-band $\mathrm{R}_{25}$ vs.~distance for CALIFA galaxies. The wedge-shaped region of the plane occupied by the sample is due to the size selection. Black circles show the full CALIFA DR3 spectroscopic sample of 528 galaxies, while blue points show the sub-sample we select for analysis in this work (100 galaxies, the same for the other panels, see \autoref{subsec:dqr}). The red diamonds represent the eight galaxies in \citet[][the same for the other panels]{Kreckel20}. Middle: $\mathrm{R}_{25}$ vs.~stellar mass and their distributions. Number distributions are normalised so that the maximum value in each panel is unity (the same for the other panels). Right: star formation rate vs.~stellar mass and their distributions.}
\label{fig:basic}
\end{figure*}

The first step in our project is to select an IFS survey suitable for analysis. As discussed above, the candidates for which public data are available are, e.g., SAMI, MaNGA, CALIFA, and MAD. The choice between them involves a tradeoff between resolution, sample size, and field of view (FoV). Of these candidates, SAMI and MaNGA have the largest samples, but comparatively lower resolution and coverage. Quantitatively, the KT18 models predicts that metallilcities in galaxies should be correlated on scales of $\sim 0.1-1$ kpc, and this prediction is consistent with the results for the eight PHANGS-MUSE galaxies analysed by \citet{Kreckel20}. Unfortunately the typical resolutions available in SAMI and MaNGA are insufficient to resolve such scales. In contrast, local MUSE surveys (e.g., MAD) have excellent spatial resolution, but their typically smaller sample sizes make them unsuitable for this parameter study. We adopt CALIFA because it has the right balance of resolution and size, achieving $\lesssim400$ pc spatial resolution for hundreds of galaxies with FoV $>1$arcmin$^2$. CALIFA aims to observe a sample of over 900 galaxies in the local universe using 250 observing nights. To date 528 galaxies have been spectroscopically observed. For these galaxies, we use the DR3 Pipe3D emission line maps produced using the moment analysis for faint lines \citep{Sanchez16a}.

For the purposes of our analysis below, we will need to characterise CALIFA's spatial resolution and FoV. The former is determined by the properties of the CALIFA PMAS/PPAK integral field spectrophotometer. The PPAK aperture broadens the point spread function (PSF) to $2\farcs 50\pm0\farcs 34$ (FWHM), substantially larger than the seeing limit \citep[][]{Sanchez16b}. The corresponding beam size projected onto our target galaxies ranges from $\sim100$ pc to $\sim600$ pc. CALIFA selects targets such that the size of the galaxy is well-matched with the instrumental FoV \citep{Walcher14}. This is advantageous, resulting in coverage over relatively similar portions of their discs, encompassing typically 2-3 effective radii. However, this selection implies that our spatial resolution will be correlated with the physical size, and thus physical properties, of the galaxy. We illustrate this effect in the left panel of \autoref{fig:basic}.

To measure metallicity from the line maps, we adopt the [N~\textsc{ii}]/[O~\textsc{ii}] metallicity diagnostic from \citet{K19a} (abbreviated as K19N2O2) as the default diagnostic of our work. K19N2O2 is one of the most reliable metallicity diagnostics in the optical spectrum, with little dependence on ionization parameter, and only marginal dependence on the ISM pressure for 4 $\leq \log[P/k_B/(\mbox{K cm}^{-3})] \leq$ 8. In practice we adopt $\log(P/k)=5$ and ionization parameter $\log(U)=-3$ throughout this work. Below we also show that we obtain fairly similar results using alternative diagnostics (see \autoref{subsec:tests}). Line ratios are corrected for dust attenuation using the attenuation curve proposed by \citet{Cardelli89}. The intrinsic error \citep[2.65\% root-mean-square error,][]{K19a} of K19N2O2 has been added and propagated through the line fluxes.

We also mask pixels in the maps based on three criteria. First we mask pixels that are potentially affected by contamination from active galactic nuclei (AGN), since standard diagnostics do not yield reliable metallicity estimates for them; we flag these pixels using the Kewley line AGN criterion \citep{Kewley01}. Our qualitative results are not significantly influenced by changing the AGN criterion (see \aref{app:agn}). Second we mask pixels where the equivalent width (EW) of H\textsc{$\alpha$} is $<6$ \AA~ \citep{Espinosa-Ponce20}, because these may reflect contamination from diffuse ionised gas rather than star-forming regions. Third we mask pixels where the signal-to-noise ratio (S/N) derived using the line flux and error maps output by Pipe3D is $<2$. We remove from our sample all galaxies containing $<500$ non-masked pixels, and this cut reduces the remaining sample to 100 galaxies. We select the minimum S/N and pixel count experimentally, and discuss our choices of thresholds in detail in \autoref{subsec:dqr}. The physical properties of the 100 galaxies that comprise our remaining sample are shown in \autoref{fig:basic}. This figure shows that our sample is representative of the full CALIFA sample in the $M_*$-$\mathrm{R}_{25}$ plane. In the $M_*$-$\mbox{SFR}$ plane our sample is representative of the part of the CALIFA sample that lies along the star-forming main sequence. We note that the full CALIFA sample also includes passive and transitional galaxies, which are excluded from our sample on the basis that they have too few pixels containing line emission bright enough to pass our S/N threshold.

\section{Analysis Method}
\label{sec:methodology}

We next describe the analysis pipeline that we apply to every galaxy. \autoref{subsec:fluc} and \autoref{subsec:2p_corr} describe the process by which we compute a galaxy metallicity fluctuation map and its two-point correlation function. \autoref{subsec:MCMC} describes our procedure for fitting a model for correlation statistics to the data, and \autoref{subsec:dqr} describes how we downselect from the full CALIFA DR3 spectroscopic sample to ensure that we limit the sample to galaxies for which we can derive reliable results.

\subsection{Metallicity fluctuation maps}
\label{subsec:fluc}

\begin{figure*}
\includegraphics[width=1.0\linewidth]{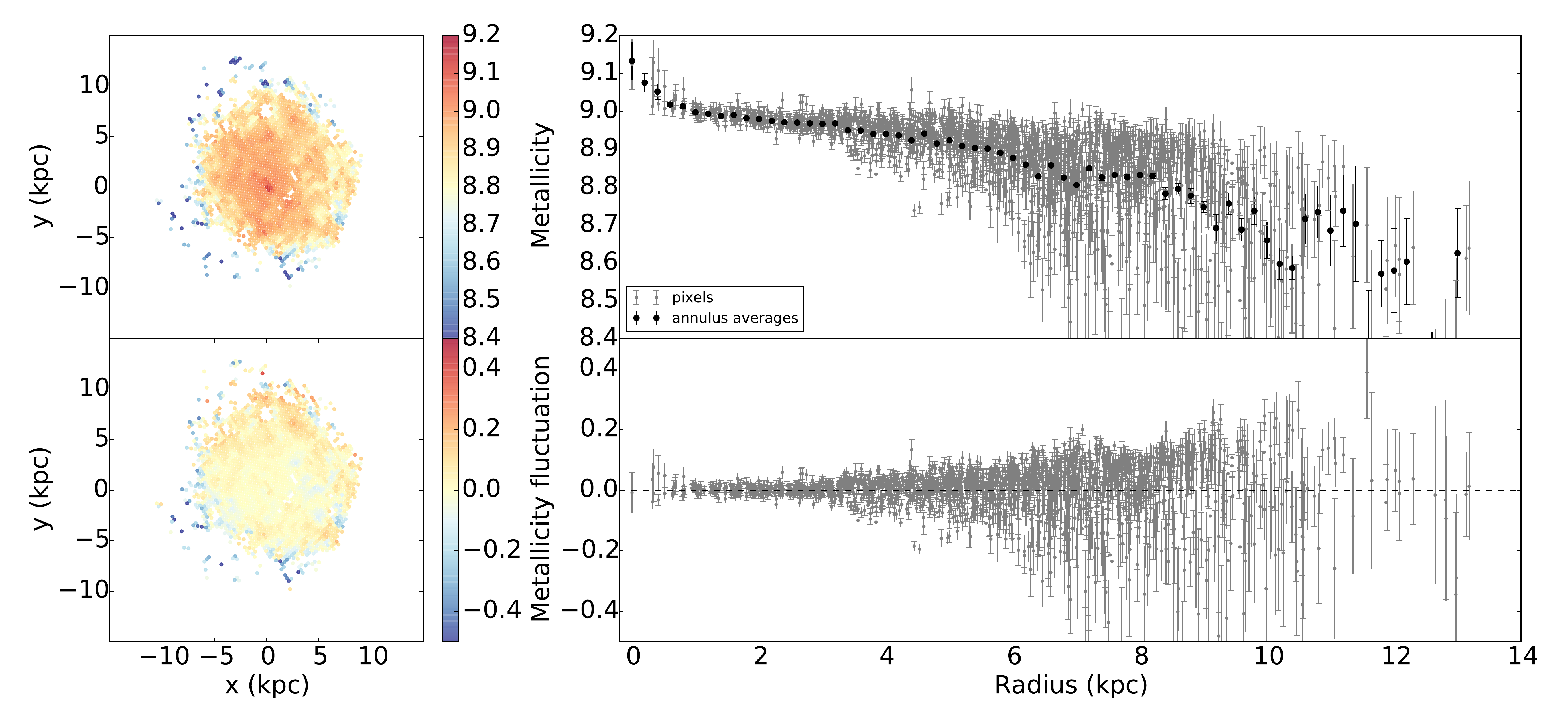}
\caption{Illustration of our procedure for generating metallicity fluctuation maps. The upper left panel shows the deprojected metallicity map of NGC0873. The upper right panel shows metallicity as a function of radius. In this panel, each grey point is an individual pixel in the metallicity map, with a corresponding error bar, and black points show means and uncertainties computed in 0.2 kpc bins. After subtracting the mean value in each annular bin, we obtain the metallicity fluctuation maps shown in the lower left panel, and the fluctuation as a function of radius shown in the lower right panel.}
\label{fig:fluc}
\end{figure*}

The first step in our pipeline is to deproject and rotate our galaxy images to produce a modified image where the galaxy has a circular profile, with the major axis of the original image aligned to the $x$ axis of the modified image. We do this by mapping the position of every pixel $(x,y)$ in the original image to a new position $(x',y')$, given by
\begin{equation}
\left[\begin{array}{c}
x'\\
y'
\end{array}\right]
=
\left[\begin{array}{cc}
\cos\theta & \sin\theta \\
-\sin\theta/\cos i & \cos\theta/\cos i
\end{array}\right]
\left[\begin{array}{c}
x\\
y
\end{array}\right],
\end{equation}
where $\theta$ is the position angle (PA) and $i$ is the inclination angle. We determine the latter using the classical Hubble formula \citep[][]{Hubble26}
\begin{equation}
\cos^2 i = \frac{(b/a)^2-q_0^2}{1-q_0^2},
\label{equ:cosi}
\end{equation}
where $b/a$ is the axis ratio in the original image and $q_0 = 0.13$ ($i=90^{\circ}$ for $b/a<q_0$). The position angle $\theta$ and the axis ratio $b/a$ (computed from eccentricity, defined as $\sqrt{1 - (b/a)^2}$) are the values produced by most recent version of the CALIFA image pipeline \citep[][see \aref{app:decom}]{Lopez19}. This pipeline imposes maximum eccentricity of 0.949 (corresponding to $b/a>0.316$, $i<85^{\circ}$), because any deprojection of images that are too close to edge-on will be extremely unreliable. For this reason, we exclude from our sample those galaxies for which the CALIFA pipeline assigns this maximum eccentricity.

Our next step is to remove the radial metallicity gradient from the deprojected map to produce a fluctuation map. We do not do this by fitting a linear function; instead, we choose to subtract the mean value, $\overline{Z_r}$, computed in annular bins of width 0.2 kpc. Our qualitative results do not depend on the choice of bin size (see \aref{app:adp}). \autoref{fig:fluc} shows an example of this procedure as applied to the galaxy NGC0873.
Our approach of subtracting bin-by-bin rather than subtracting a linear fit is a safety measure to guard against false-positive signals for metallicity correlation that might arise due to non-linearities in the metallicity diagnostic. Specifically, suppose that our metallicity diagnostic is not perfectly linear in the input metallicity, i.e., when applied to a patch of a galaxy with true (log) metallicity $Z$, the estimate returned by the diagnostic is $Z_{\rm est} = Z + f_{\rm nl}(Z)$, where $f_{\rm nl}(Z)$ is some small but non-linear function. Then even if were to examine a galaxy where the true metallicity is an exactly linear function of radius, the estimated metallicity as a function of radius would not be an exactly linear function due to the error $f_{\rm nl}$, and subtracting a linear gradient from it would leave behind non-trivial structure that would be entirely due to $f_{\rm nl}$. Calculation of the spatial correlation of the resulting map might then yield a signal dominated by this effect. Subtracting the mean metallicity computed independently in each annular bin prevents this problem from occurring, at the possible cost of reducing the true correlation signal by masking real correlations in the radial direction. 

The uncertainty of the mean metallicity in each annular bin is computed using bootstrapping. Our procedure is as follows: for each pixel $i$ in the map, we have an estimate of the metallicity $Z_i$ and an uncertainty $\sigma_{Z,i}$. In each bootstrap trial, we generate a random realisation of the metallicity map by drawing a random metallicity for each pixel from a Gaussian distribution with mean $Z_i$ and standard deviation $\sigma_{Z,i}$. We then compute the annular mean metallicities for this random map. We repeat this procedure 20 times, thereby deriving a sample of 20 values for the mean metallicity in each annular bin. We take the mean and standard deviation of these 20 realisations as our central value $\overline{Z_r}$ and $1\sigma$ uncertainty $\sigma_{\overline{Z_r}}$. Note that the uncertainty of any individual pixel will change after subtracting the mean value in its annular bin. The metallicity fluctuation $Z'_i$ and its uncertainty $\sigma_{Z', i}$ for each pixel are then
\begin{eqnarray}
Z'_i & = & Z_i - \overline{Z_r}\\
\sigma_{Z', i}^2 & = & \sigma_{Z, i}^2 + \sigma_{\overline{Z_r}}^2.
\end{eqnarray}

\subsection{Two-point correlations}
\label{subsec:2p_corr}

\begin{figure}
\includegraphics[width=1.0\linewidth]{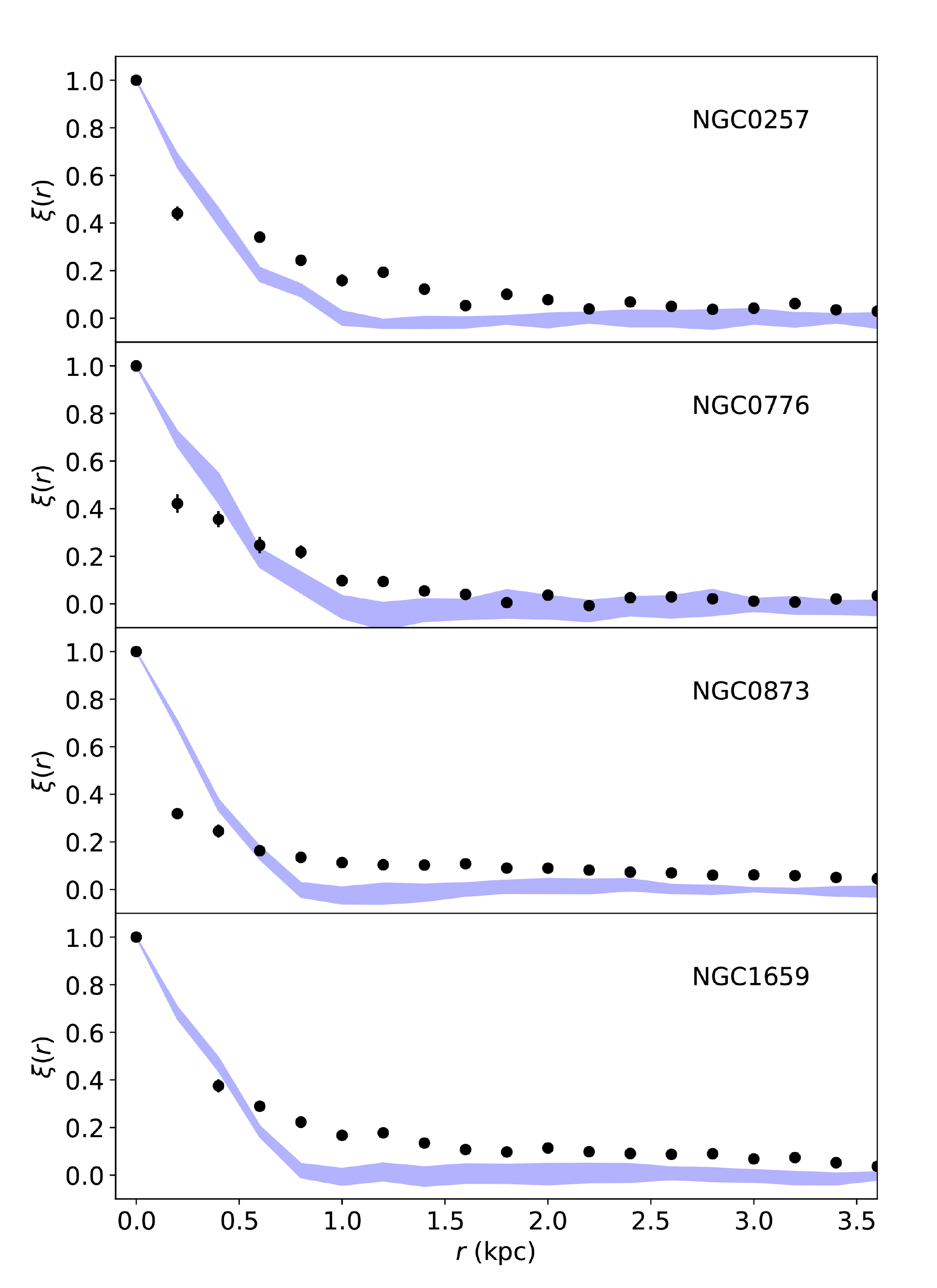}
\caption{The two-point correlation functions (the black dots and black error bars) of the four example galaxies, and the comparison with the two-point correlation function of blue noise (the blue band). The black dots above the band are considered to be the genuine signals.}
\label{fig:TPCF}
\end{figure}

The result of the procedure we have just described is a metallicity fluctuation map, which has (by construction) zero mean. We now compute the two-point correlation function of this map, which, for a zero-mean map, is given by
\begin{equation}
\label{equ:TPCF}
    \xi(\mathbf{r}) = \frac{\left\langle Z'(\mathbf{r} + \mathbf{\mathbf{r}'}) Z'(\mathbf{r}') \right\rangle}{\left\langle Z'(\mathbf{r}')^2\right\rangle},
\end{equation}
where $Z'(\mathbf{x})$ is the metallicity fluctuation (i.e., the quantity we obtain after subtracting the mean metallicity in annular bins) at position $\mathbf{x}$ in the map, and the angle brackets $\left\langle\cdot\right\rangle$ denote averaging over the dummy position variable $\mathbf{r}'$. Clearly $\xi(\mathbf{r})$ is bounded in the range $-1$ to $1$. In practice we are interested only in the two-point correlation as a function of scalar separation $r$ rather than vector separation $\mathbf{r}$. We compute this by taking every pair of pixels $(i,j)$ in the deprojected, mean-subtracted map, computing their separation $r_{ij}$, and binning the pixel pairs by $r_{ij}$. We then compute the two-point correlation function for the $n$th bin, spanning the range of separations $(r_n, r_{n+1})$ as
\begin{equation}
    \xi_n = \frac{\sigma_{Z'}^{-2}}{N_n} \sum_{r_n < r_{ij} \leq r_{n+1}} Z'_i Z'_j,
    \label{eq:xi_definition}
\end{equation}
where the sum runs over the $N_n$ pixel pairs $(i,j)$ for which $r_n < r_{ij} \leq r_{n+1}$, and
\begin{equation}
    \sigma_{Z'}^2 = \frac{1}{N_p} \sum_{i=1}^{N_p} {Z'_i}^2
\end{equation}
is the total variance of the $N_p$ pixels in the map\footnote{Note that, because our map is finite and non-periodic, the definition of the correlation function becomes ambiguous for separations comparable to the size of the map. The computational manifestation of this is that, while the number of pairs $N_n$ in a given radial separation bin is always larger than the number of map pixels $N_p$ for separations much smaller than the map size, for sufficiently large separations this ceases to be the case, and then in principle \autoref{eq:xi_definition} could return a value $\xi_n<-1$ or $>1$, which is unphysical. To avoid this complication, we only compute the correlation function for separations up to 4.0 kpc, corresponding to 20 bins for our 0.2 kpc bin width. This is large enough that few galaxies have significant correlations on this scale, and thus we lose no information by making this truncation.}. We determine the uncertainty of the derived two-point correlation using the same bootstrapping procedure described above, i.e., we generate 20 random realisations of the metallacity fluctuation map by drawing a value for each pixel from a Gaussian with mean $Z'_i$ and uncertainty $\sigma_{Z',i}$. We then compute the two-point correlation function of this random map, and record the mean (as the central value) and standard deviation (as the $1\sigma$ uncertainty) of the value of the correlation function in every separation bin. We show the two-point correlation function we derive via this procedure for four galaxies (NGC0257, NGC0776, NGC0873, and NGC1659) in \autoref{fig:TPCF}. Note that error bars are shown in the figure, but are in many cases too small to be easily visible. The errors are small because the correlation function involves a sum over a large number of pixel pairs, which leads to the errors being averaged-down considerably.

\autoref{fig:TPCF} shows clear evidence of spatial correlation in the metallicity distribution. However, we expect to find some correlation in the map simply as a result of the combination of beam smearing and masking of pixels (both those with AGN contamination, and those where the CALIFA data contain insufficient signal to allow derivation of a metallicity). It is therefore important to verify that the correlation we see is in excess of what  might be expected due to these effects. We check for this possibility as follows: for each galaxy, we construct a pure, uncorrelated noise map with a resolution five times higher than the resolution of the real map, and set the metallicity fluctuation in each pixel to a Gaussian random value with zero mean and unit variance (though the value of the variance has no effect on the correlation). We then convolve the noise map with a Gaussian beam with the same FWHM as the estimated CALIFA PSF for that galaxy\footnote{PSF values are provided in \cite{Sanchez16b}. In principle the noise should correlated, since CALIFA's dithering strategy mixes together the light from different fibres into a single pixel \citep{Husemann13}. This effect is minimal for this analysis because we are only concerned with the correlations on larger scales. On such scales, there is minimal covariance between pixels.}, rebin the resulting smeared map down to the same pixel size as the actual CALIFA map, and apply the same pixels mask as for the real map (i.e., if a pixel is masked in the real map, we mask the corresponding pixel in the noise map). We then compute the two-point correlation function of the noise map using the same pipeline we apply to the real map. We repeat this procedure 10 times for each galaxy and measure the mean and standard deviation of the resulting correlations. We show the correlation functions of the noise maps as blue bands in \autoref{fig:TPCF}. It is clear that, for the galaxies shown (and for the vast majority of the sample), the correlation measured in the real map substantially exceeds the correlation of the noise map for all but the smallest separation bins.\footnote{Careful readers may notice that the blue noise maps are actually \textit{more} correlated than the real measurements at separations $r\lesssim 0.5$ kpc. The reason for this is that, in constructing the noise maps, we have not added any observational errors beyond beam smearing. In contrast, the real observations are affected by both beam smearing and errors in recovering the metallicity -- both from photometric uncertainties in the line fluxes, and from imperfections in the metallicity diagnostics used to transform line fluxes into metallicities. As we show in \autoref{subsec:MCMC}, the effect of these errors is, at least approximately, to reduce the two-point correlation by a constant, separation-independent factor $f$ at all separations larger than one pixel. Thus observational errors in the real maps can yield a measured correlation function that is smaller than the correlation function of a map that consists of nothing but beam-smeared noise, but is without observational errors. Conversely, if we added observational errors to the blue noise maps, the result would be to lower the correlation in all bins except the first, while leaving the shape unchanged; for a reasonable choice of noise level, this would render the correlations in the nosie maps smaller than those in the real observations at all separations.}

\begin{figure}
\includegraphics[width=1.0\linewidth]{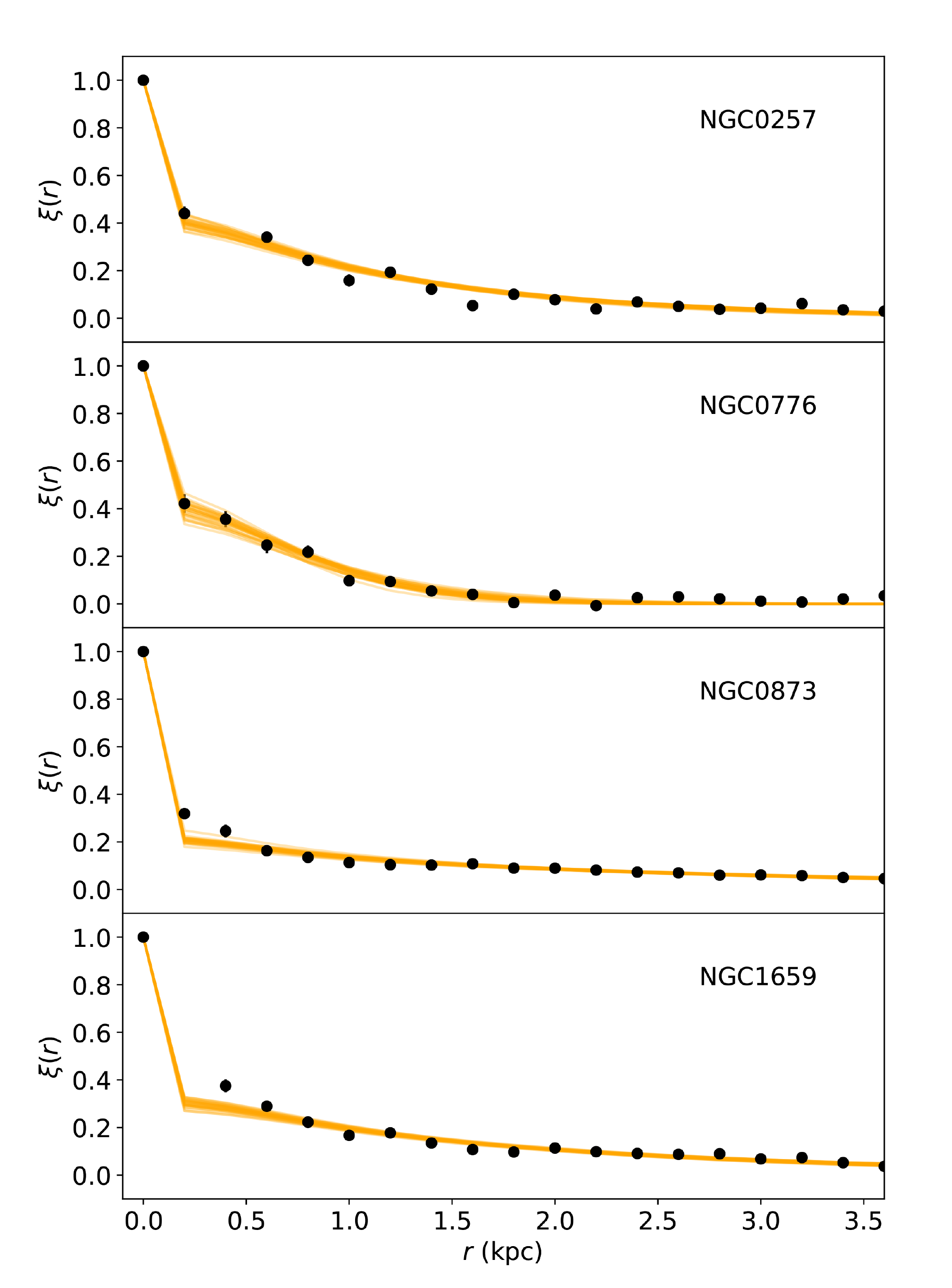}
\caption{Comparison between MCMC-fit models for the two-point correlation function (\autoref{equ:model}; light orange curves) of the four example galaxies and observed two-point correlation functions (points with error bars; same as in \autoref{fig:TPCF}). The model fits shown represent 30 samples randomly selected from the converged MCMC chains. The break point at 0.2 kpc (the first bin edge) is a result of the perfect autocorrelation of errors within a single pixel -- see main text.}
\label{fig:curves}
\end{figure}

\subsection{Parametric model fit}
\label{subsec:MCMC}

In addition to measuring the two-point correlation function, we carry out a parametric fit, using the functional form proposed by KT18. The model is based on the idea that the metallicity fluctuation statistics in a galaxy are controlled by a competition between metal injection events, which increase the correlation, and diffusion of metals through the ISM, which decreases it. We modify the model to account for the effects of beam smearing and noise in \aref{app:kt18_generalisation}. The prediction derived there (\autoref{equ:corr_conv_noise}), which we repeat here for convenience, is
\begin{eqnarray}
\lefteqn{\xi_{\rm model}(r) = \frac{2}{\ln\left(1 + \frac{2\kappa t_*}{\sigma_0^2/2}\right)
} \left[\frac{\Theta(r-\ell_{\rm pix})}{f} + \Theta(\ell_{\rm pix}-r)\right]
}
\nonumber \\
& &
\int_0^\infty e^{-\sigma_0^2 a^2/2} \left(1 - e^{-2 \kappa t_* a^2}\right) \frac{J_0(ar)}{a} \, da,
\label{equ:model}
\end{eqnarray}
where $\sigma_0^2 = \sigma_{\rm beam}^2 + 2 \sigma_{\rm inj}^2$, $\sigma_{\rm beam}$ is the dispersion of the observational beam in linear (as opposed to angular) distance, $\sigma_{\rm inj}$ is the physical width over which supernovae inject metals, $\kappa$ is the diffusion coefficient, $t_*$ is the star formation duration, $\ell_{\rm pix}$ is the size of a pixel in the observed map, and $f$ is the factor by which observational errors in the derived metallicities increase the variance in the metallicity fluctuations compared to the true variance. Here $\Theta(x)$ is the Heaviside step function, and the purpose of the term in square brackets containing $\Theta(x)$ terms is to account for the fact that errors are perfectly correlated within a single pixel of the observed image, and completely uncorrelated (at least in our approximation) between different pixels -- see \aref{app:kt18_generalisation} for details.

\begin{figure}
\includegraphics[width=1.0\linewidth]{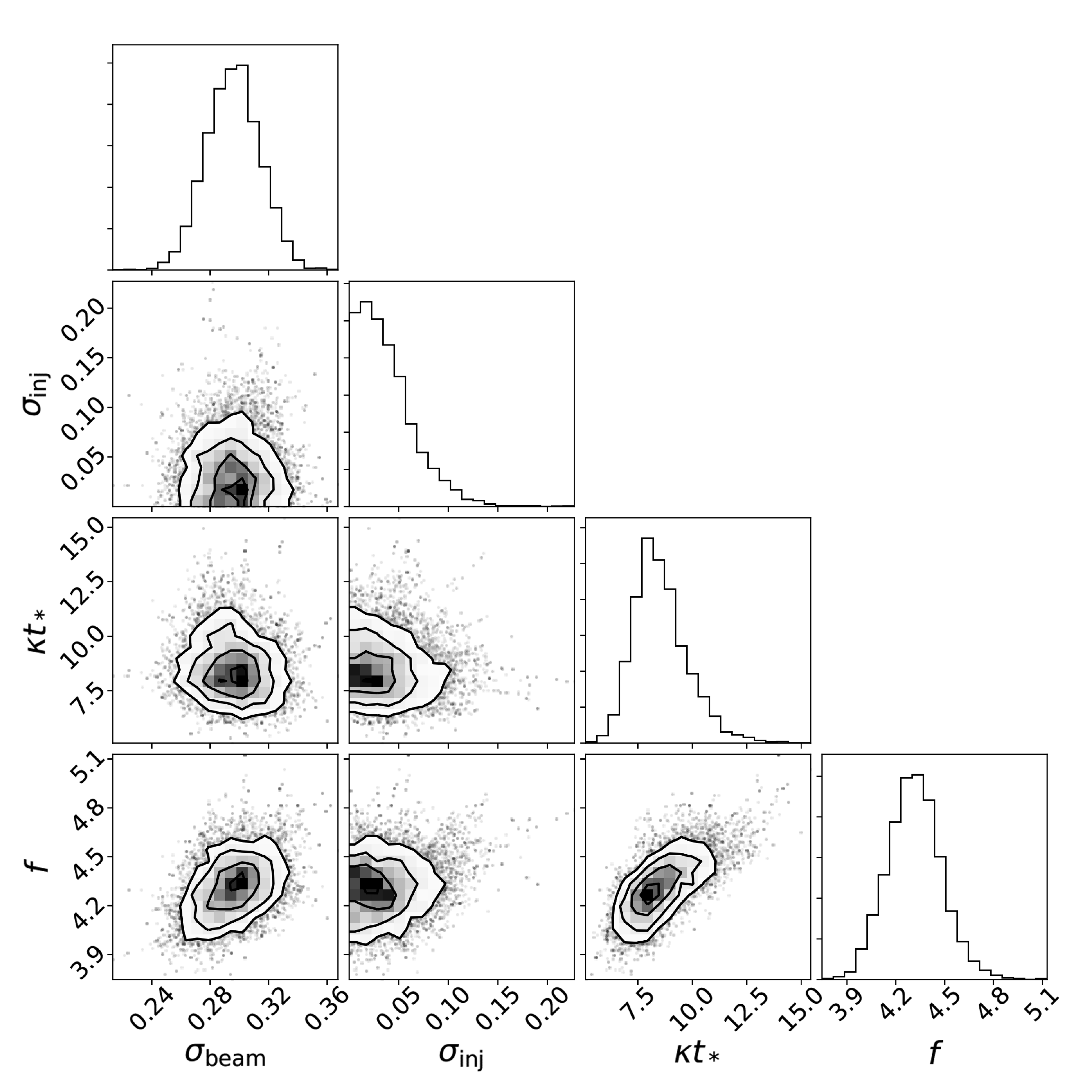}
\caption{The distribution of our four fitting parameters describing the two-point correlation function measured in NGC0873, as derived from our MCMC. In this plot, $\sigma_{\rm beam}$ and $\sigma_{\rm inj}$ have units of kpc, and $\kappa t_*$ has units of kpc$^2$. In each panel, the heat map colours black, dark grey, light grey, and white (i.e. the four contours from inside to outside) show probability densities corresponding to 1, 2, 3, and $4\sigma$, respectively. Partially-transparent points show individual samples in regions of lower probability.}
\label{fig:par}
\end{figure}

We fit the measured correlation function for each galaxy in our sample to the functional form given by \autoref{equ:model} using the Python package \textsc{emcee} \citep{emcee}, an implementation of an affine-invariant ensemble sampler for Markov chain Monte Carlo (MCMC). We fit four parameters: $\sigma_{\rm beam}$, $\sigma_{\rm inj}$, $\kappa t_*$, and $f$. The likelihood function is given as
\begin{eqnarray}
\lefteqn{\ln p(\xi \mid \sigma_{\rm beam}, \sigma_{\rm inj}, \kappa t_*, f) =
}
\nonumber \\
& &
-\frac{1}{2} \sum_n \left[\frac{(\xi_{\rm model} - \xi_{\rm obs})^2}{\sigma_{\xi, \rm obs}^2} + \ln(\sigma_{\xi, \rm obs}^2)  \right],
\end{eqnarray}
where quantities subscripted `model' are evaluated from \autoref{equ:model}, quantities subscripted `obs' refer to the correlation function measured in the observations, and the sum is over the $n$ bins for which we have evaluated the observed correlation function.

As for the prior, all four parameters are physically required to be positive definite. For $\sigma_{\rm inj}$, $\kappa t_*$, $f$ we therefore adopt flat priors for all positive values, and priors of zero for negative values. For $\sigma_{\rm beam}$, we choose our prior based on the measured distribution of PSFs for CALIFA: for each galaxy, an estimated FWHM of PSF is provided by \citet{Sanchez16b}. We assume a Gaussian distribution of PSF with the estimated PSF as the mean and $0\farcs15$ as the standard deviation, and convert into a physical length given the distance to each galaxy. We also set the prior on $f$ to $1/f$, i.e. flat in logarithmic space. We carry out the MCMC fit using 100 walkers, run for 500 steps in total; visual examination shows that the chains are well-converged after $\sim 250$ steps, so we take the first 350 steps as a burn-in period for safety, and derive the posterior PDF from the final 150 steps.

In order to evaluate whether the functional form given by \autoref{equ:model} is a reasonable match to the observations, in \autoref{fig:curves} we randomly select 30 models from our total of 15,000 post-burn-in samples (150 steps $\times$ 100 walkers) and plot these models together with the observed two-point correlations for several example galaxies. The curves produced by the best fits pass through the observed two-point correlations well, meaning that the model is capable of providing a good description of the data.  \autoref{fig:par} shows the distributions of the parameters for the example galaxy NGC0873 obtained from our MCMC fit; while exact parameter values of course vary between galaxies, this example is representative of the qualitative trends in our fits. A few remarks are in order. First, we obtain only an upper limit for $\sigma_{\rm inj}$. This is not surprising: this quantity affects the shape of the correlation only on scales comparable to $\sigma_{\rm inj}$, and in a beam-convolved map, it appears only added in quadrature with $\sigma_{\rm beam}$. KT18's modelling suggests that the true value of $\sigma_{\rm inj}$ should be $<100$ pc, which is always smaller than the CALIFA beam. Consequently, given our resolution, we can only constrain the injection width to be smaller than the CALIFA beam size. By contrast, we obtain strong constraints on $\kappa t_*$, because this quantity determines the shape of the correlation function on large scales that are well-resolved by the CALIFA beam. In what follows, we will take the physical quantity correlation length
\begin{equation}
    \label{equ:lcorr_1}
    l_{\rm corr} \equiv \sqrt{\kappa t_*},
\end{equation}
to which we refer as the correlation length, as the main quantity of interest. A final note is that our best-fitting value of $f$ is $\sim4.3$, which is typical of most of our fits. This means that the uncertainty in the metallicities of individual pixels is increasing the apparent dispersion in total metallicity by a factor of 4.3, which corresponds to the single-pixel metallicity uncertainty being a factor of $\sim2$ larger than the intrinsic metallicity dispersion of the galactic disc. Given what we know about the uncertainties of metallicity indicators, this is a plausible figure. 

\begin{figure}
\includegraphics[width=1.0\linewidth]{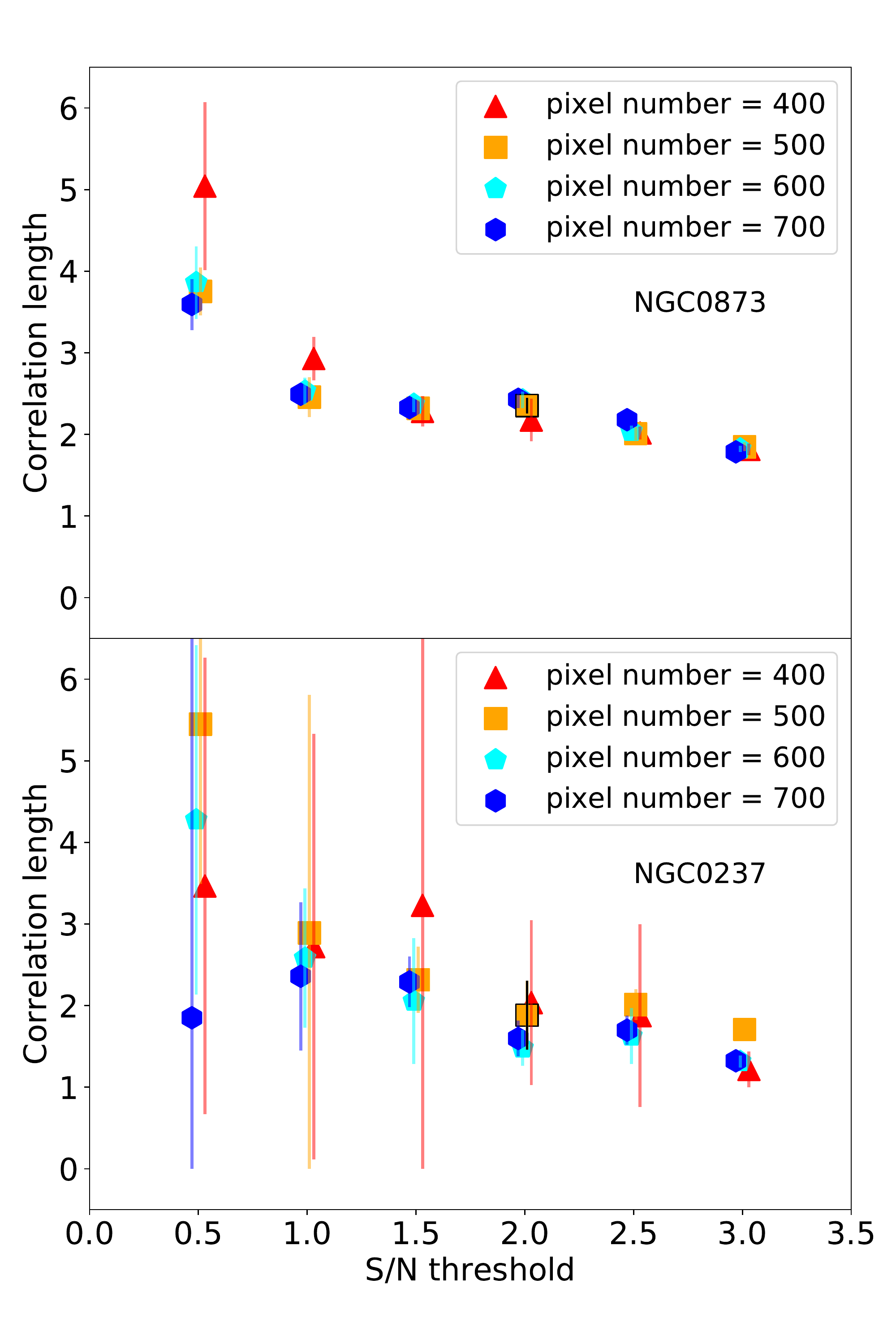}
\caption{Estimated value of correlation length as a function of S/N threshold and pixel number for  NGC0873 and NGC0237. Points and error bars show the mean and dispersion of 50 realisations; see main text for details. Based on this anslysis, we adopt a data quality requirement of 500 pixels above a threshold S/N $= 2.0$; this point is marked with black edges.}
\label{fig:thres}
\end{figure}

\subsection{Data quality requirements}
\label{subsec:dqr}

Before the we can apply the analysis method described above the full CALIFA sample, we must select galaxies with suitable data quality. The two-point correlation function is defined in terms of a sum over pixel pairs. This means that if a map contains too few pixels, the result will be a noisy two-point correlation function, potentially lacking information on large spatial scales. On the other hand, if a map contains a large number of pixels for which the metallicity is very uncertain, these pixels will also reduce the robustness of the measured two-point correlations. These two considerations mean that, while we wish to apply a minimum S/N ratio requirement to our maps and mask pixels that fall below it, there is a trade-off between accuracy of line flux of single pixels and the number of pixels of the whole galaxy. A higher S/N threshold yields more pixel-level accuracy at the price of worse statistical accuracy in evaluation of the two-point correlation function, while a lower S/N threshold yields the opposite.

In order to optimise this trade-off, we investigate two example galaxies, NGC0873 and NGC0237. We select these two because they have a large number of high-S/N pixels and thus yield high-quality results; we can then artificially degrade the maps to lower quality, in order to determine at what point our results become unreliable. To carry out this test, we consider six possible S/N thresholds for the line fluxes: 0.5, 1.0, 1.5, 2.0, 2.5, and 3.0. We also consider four different minimum pixel counts: 400, 500, 600, and 700. For each possible combination of S/N threshold and pixel number, we carry out the following steps: (1) we mask all pixels that fall below the specified S/N threshold; (2) we randomly mask enough additional pixels reduce the total number remaining to our target pixel count; (3) we apply our parametric fitting procedure (\autoref{subsec:MCMC}) to the resulting map and obtain the 50th percentile value of $l_{\rm corr}$, the primary quantity of interest for the remainder of our analysis. We repeat this procedure 50 times for each combination of S/N and pixel number. \autoref{fig:thres} shows how $l_{\rm corr}$ changes with S/N threshold and minimum pixel number; points and error bars indicate the means and standard errors of the 50 realisations. It is clear that, as one might expect, the 50th percentile estimate of $l_{\rm corr}$ fluctuates more in the experiments with fewer pixels. The S/N threshold has a large effect if we choose a very weak threshold of 1, but there is little difference in the results derived with thresholds $\geq 2$. Based on the results shown in this figure, we adopt a minimum S/N ratio at 2 and a minimum of 500 pixels above this threshold. The full CALIFA sample contains 100 galaxies that satisfy this condition, and we limit our analysis to these 100 galaxies from this point forward.

\begin{table*}
\caption{Global properties and correlation lengths of CALIFA galaxies. Columns are as follows: (1) galaxy name; (2) position angle; (3) ellipticity; (4) inclination angle; (5) distance; (6) $r$-band $\mathrm{R}_{25}$ (7) FWHM of PSF; (8) log stellar mass; (9) H\textsc{$\alpha$} star formation rate; (10) - (12) correlation length derived using the K19N2O2, PPN2, and PPO3N2 diagnostics, respectively -- the central value is the 50th percentile, and the error bars show the 16th to 84th percentile range. The position angle and ellipticity are from \citet{Lopez19}, and the stellar mass column is from \citet{Lacerda20}. We derive distances from the redshifts provided in the \href{https://califaserv.caha.es/CALIFA_WEB/public_html/?q=content/dr3-tables\#BASIC}{basic information table} on the CALIFA official website. The SFR values are derived from dust-corrected H\textsc{$\alpha$} \citep{Sanchez20b}. This table is a stub to show structure; the full table is available in the electronic publication.}
\label{tbl:basic}
\begin{tabular}{cccccccccccc}
\hline
Name & PA & $\epsilon$ & $i$ & $D$ & $\mathrm{R}_{25}$ & PSF & $\log(M_*)$ & $\log(\mathrm{SFR})$ & $l_{\rm corr}$ (K19N2O2) & $l_{\rm corr}$ (PPN2) & $l_{\rm corr}$ (PPO3N2) \\
& ($^{\circ}$) &  & ($^{\circ}$) & (Mpc) & (kpc) & ($\arcsec$) & ($\mathrm{M}_{\odot}$) & (M$_{\odot}$ yr$^{-1}$) & (kpc) & (kpc) & (kpc)\\
(1) & (2) & (3) & (4) & (5) & (6) & (7) & (8) & (9) & (10) & (11) & (12)\\
\hline
UGC00005 & 136 & 0.490 & 60 & 109.5 & 12.831 & 3.17 & $11.119\pm0.085$ & $0.897\pm0.059$ & $1.436_{-0.177}^{+0.191}$ & $1.248_{-0.100}^{+0.099}$ & $0.987_{-0.079}^{+0.078}$ \\%[1.25ex]
\\
NGC7819 & 184 & 0.463 & 58 & 74.9 & 9.611 & 2.74 & $10.426\pm0.082$ & $0.392\pm0.060$ & $0.342_{-0.140}^{+0.100}$ & $0.358_{-0.072}^{+0.102}$ & $0.418_{-0.097}^{+0.069}$ \\%[1.25ex]
\\
IC1528 & 163 & 0.631 & 70 & 57.2 & 8.822 & 2.13 & $10.446\pm0.088$ & $0.205\pm0.060$ & $0.575_{-0.083}^{+0.067}$ & $0.957_{-0.037}^{+0.034}$ & $0.670_{-0.034}^{+0.032}$ \\%[1.25ex]
\hline
\end{tabular}
\end{table*}

\section{Results}
\label{sec:results}

As noted above, we can interpret the correlation length $l_{\rm corr} = \sqrt{\kappa t_*}$ as a characteristic length of ISM mixing. The main result of this work is the measurement of the correlation lengths of metallicity fluctuations in CALIFA galaxies. We report the 50th percentile and the 16th to 84th percentile range of correlation length for each galaxy in our sample in \autoref{tbl:basic}. The mean correlation length for our 100 galaxies is 0.97 kpc, while the median is 0.66 kpc.

\subsection{Reliability tests}
\label{subsec:tests}

\begin{figure}
\includegraphics[width=1.0\linewidth]{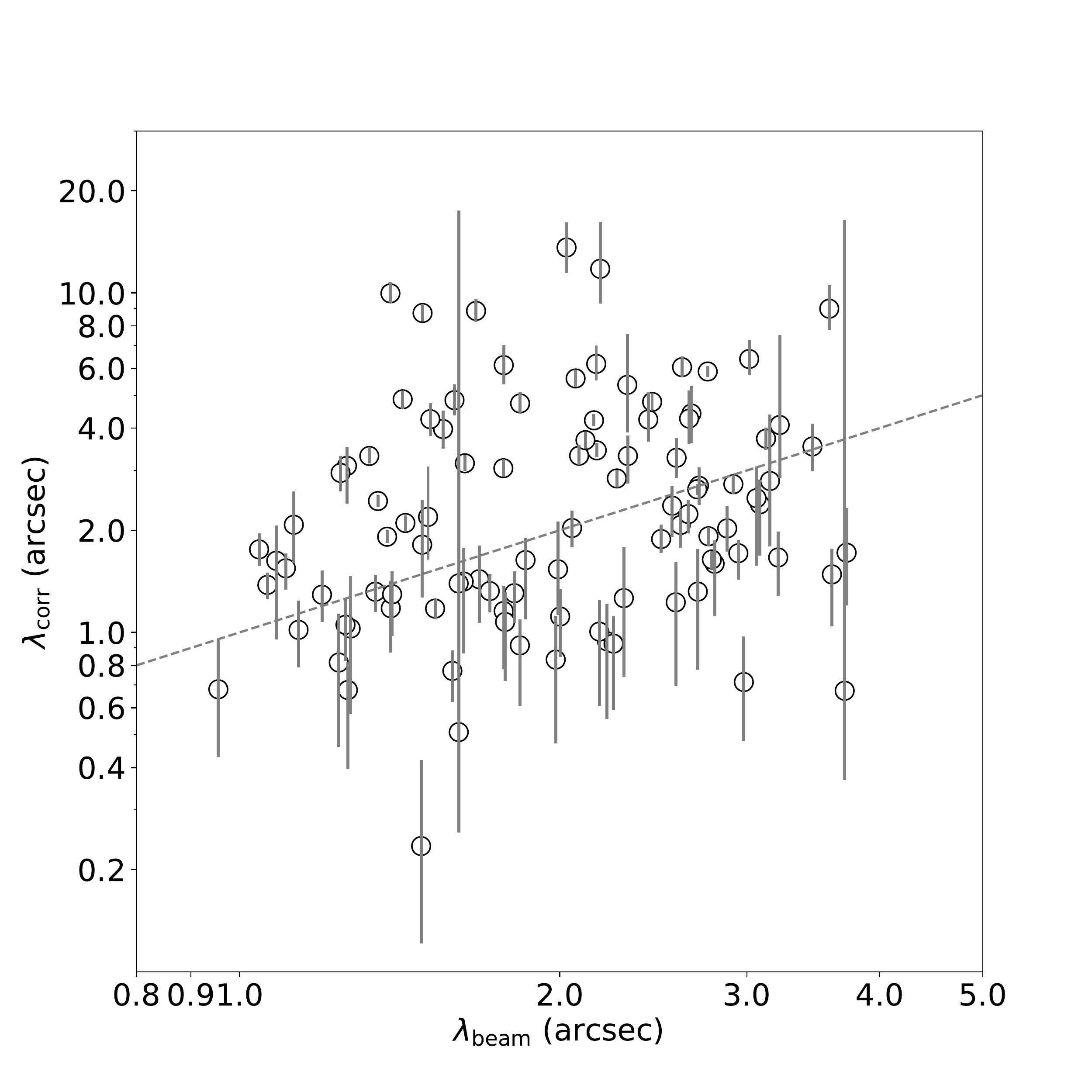}
\caption{The angular correlation length as a function of the angular semi-major axis of the telescope beam projected onto the face of the galaxy. The Pearson correlation between $\lambda_{\rm corr}$ and $\lambda_{\rm beam}$ is $0.20\pm0.07$. The black dashed line is the 1-1 line.}
\label{fig:beam}
\end{figure}

\begin{figure}
\includegraphics[width=1.0\linewidth]{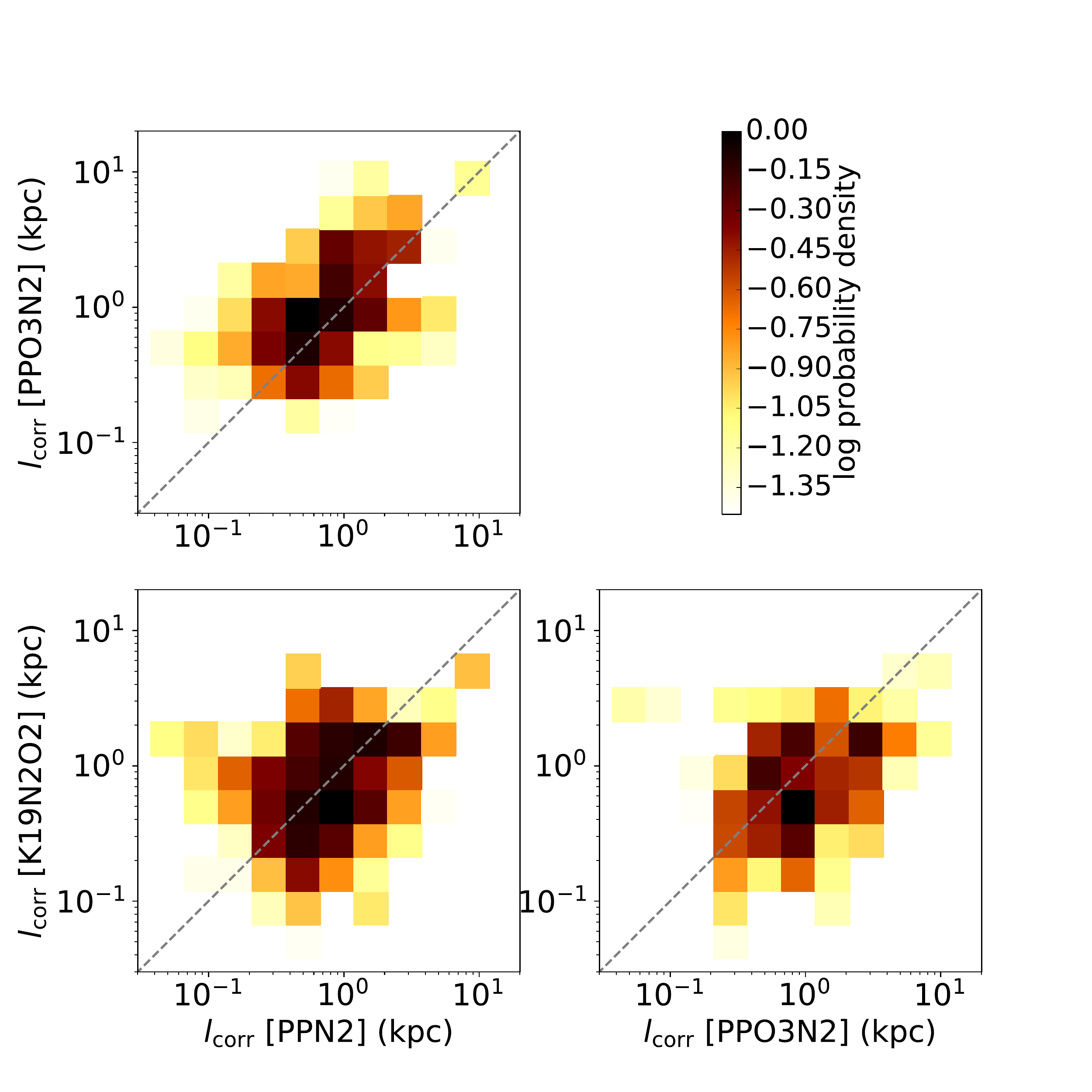}
\caption{Joint probability densities for correlation lengths using three possible pairs of diagnostics: K19N2O2 (our fiducial choice), PPN2, and PPO3N2. The colour maps show probability marginalised over the posterior PDFs of $l_{\rm corr}$ for all galaxies via the bootstrapping procedure described in the main text. Probabilities are normalised so that the maximum pixel value in each panel is unity. In this plot, perfect agreement would consist of the probability being non-zero only for pixels that lie along the diagonal 1-1 lines shown in each panel. The Pearson correlations for the distributions shown in each panel (also computed via bootstrapping) are $0.46\pm0.10$ (upper left), $0.32\pm0.08$ (lower left) and $0.36\pm0.08$ (lower right), respectively.}
\label{fig:corner}
\end{figure}

\begin{figure*}
\includegraphics[width=1.0\linewidth]{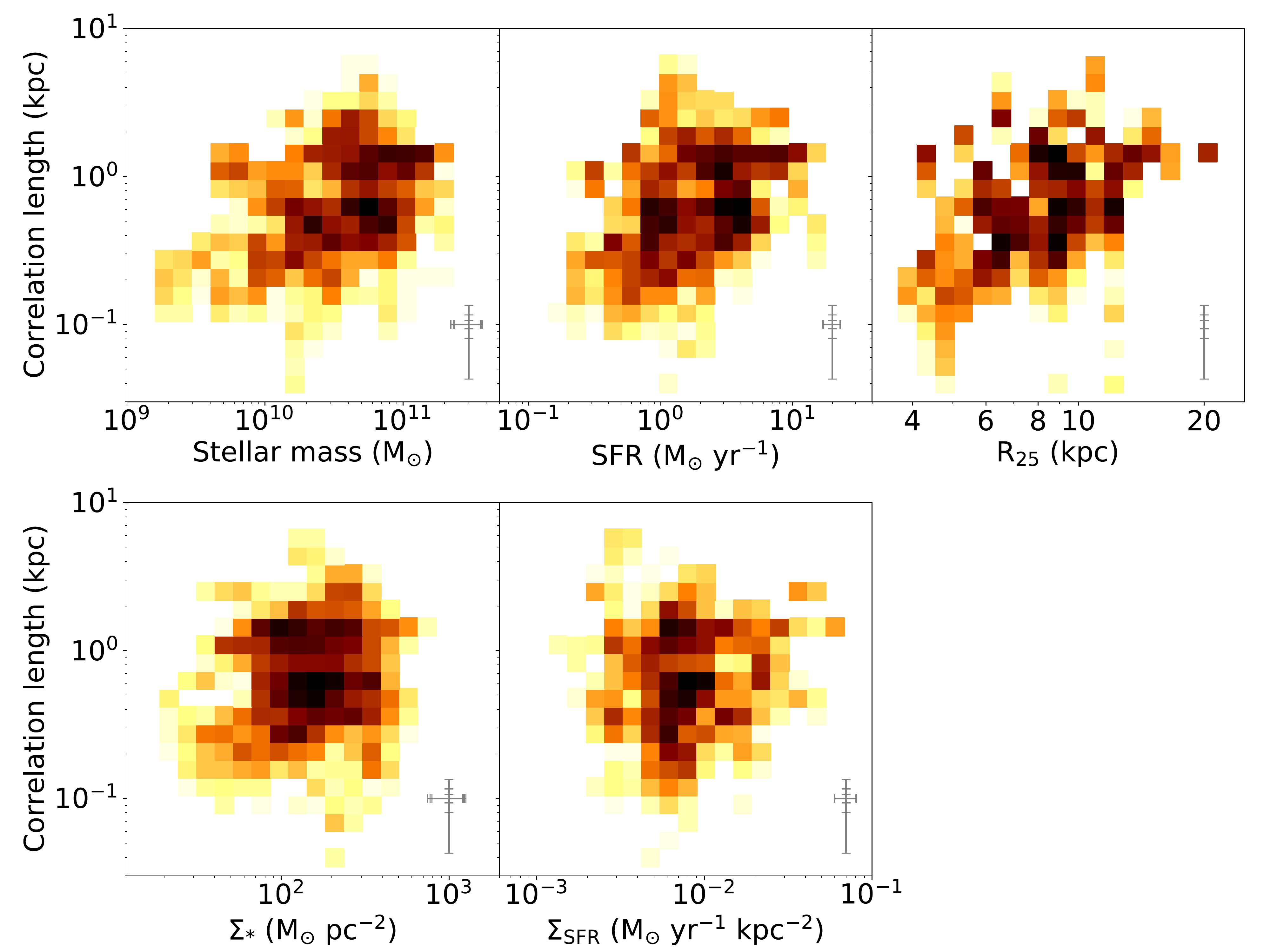}
\caption{Probability density map of correlation length vs. stellar mass (upper left), SFR (upper middle), $\mathrm{R}_{25}$ (upper right), stellar mass surface density (lower left), and SFR surface density (lower middle). The probabilities shown are marginalised over the posterior PDF of correlation length, as discussed in the main text. Probabilities are normalised so that the maximum pixel value in each panel is unity. To give a sense of the typical uncertainties on an individual galaxy, the error bars shown in the lower right corner of each panel show the 16th, 50th, and 84th percentiles of the distribution of the error intervals (proportional to errors on a logarithmic scale), i.e., 84\% of the sample has an uncertainty in correlation length smaller than the outermost set of vertical error bars, 50\% of the sample has an uncertainty smaller than the middle set, and 16\% have an uncertainty smaller than than the innermost set. The Pearson correlations for the distributions shown in each panel (also computed via bootstrapping) are $0.30\pm0.09$ (stellar mass), $0.29\pm0.06$ (SFR), $0.36\pm0.07$ ($\mathrm{R}_{25}$), $0.09\pm0.07$ ($\Sigma_*$), and $0.04\pm0.07$ ($\Sigma_{\rm SFR}$).}
\label{fig:mass_sfr}
\end{figure*}

We carry out two experiments to check the reliability of our correlation lengths. First, we seek to ensure that the correlations lengths we recover are real measurements, and are not simply reflecting the spatial resolution of the underlying maps. Our test against noise maps in \autoref{subsec:2p_corr} provides some confidence in this regard, but we carry out an additional test here, by checking whether there is correlation between the size of the telescope beam and our recovered correlation lengths. For the purposes of this test, we work in angular units to avoid introducing artificial correlations due to dependence on galaxy distance. We define the projection-corrected angular beam size to be
\begin{equation}
\label{equ:ang_beam}
\lambda_{\rm beam} = \frac{\sigma_{\rm beam,a}}{\cos i},
\end{equation}
and the angular correlation length to be
\begin{equation}
\label{equ:ang_corr}
\lambda_{\rm corr} = \frac{l_{\rm corr}}{D},
\end{equation}
where $\sigma_{\rm beam,a}$ is the angular beam size  (which has the usual relation to the angular FWHM, $\sigma_{\rm beam,a} = \mbox{FWHM}/2.355$), $\cos i$ is computed from \autoref{equ:cosi}, and $D$ is distance. The quantity $\lambda_{\rm beam}$ is the angular size of the semi-major axis of the telescope beam, projected onto the face of the galaxy, while $\lambda_{\rm corr}$ is the angular size of our recovered correlation. We show $\lambda_{\rm corr}$ as a function of $\lambda_{\rm beam}$ in \autoref{fig:beam}. We draw samples of $l_{\rm corr}$ from the posterior PDF for each galaxy, compute the Pearson correlation between $\lambda_{\rm beam}$ and $\lambda_{\rm corr}$ for one realisation, and repeat the process 50 times. This gives Pearson correlation value of $0.20\pm0.07$. It is clear that there is significant scatter in the measured $\lambda_{\rm corr}$ at a given $\lambda_{\rm beam}$, since $\lambda_{\rm beam}$ spans a factor of $<4$, while $\lambda_{\rm corr}$ varies by a factor of $\sim50$. The scattered circles clearly deviates from the 1-1 line. However, there still remains some slight correlation, which we interpret as the result of the CALIFA selection, which, as discussed in \autoref{sec:data}, favours galaxies whose effective radii are closely matched to the (distance-dependent) telescope field of view. This suggests an important design consideration for future surveys: while selecting galaxies to match the telescope FoV obviously maximises observational resources, it also induces spurious correlations between resolution and galaxy physical properties that can be difficult to disentangle.

The other check we perform is to compare our results derived using the K19N2O2 diagnostic to those derived from two other diagnostics, PPN2 and PPO3N2 \citep[][hereafter PP04]{PP04}, in order to ensure that are results are not dominated by systematic errors induced by the diagnostic. The results are shown in \autoref{fig:corner}. The three probability density subplots show the joint distributions of $l_{\rm corr}$ derived using maps produced by the three possible pairs of diagnostics. Note that this plot is not just the joint distribution of 50th percentile values. Instead, we properly integrate over the full posterior PDF of $l_{\rm corr}$ as returned by our MCMC fit via a bootstrapping procedure: we randomly select sample values of $l_{\rm corr}$ from our converged MCMC chains for each galaxy and each diagnostic, and the probability densities shown in the figure are the densities of these sample points. As illustrated in the figure, the sample points are relatively closely clustered around the 1-1 line, indicating a high degree of consistency. To quantify this, we compute the Pearson correlation coefficients between the different metallicity diagnostics. We again marginalise over the posterior PDFs by bootstrapping: we draw samples of $l_{\rm corr}$ from the posterior PDF for each galaxy and each diagnostic pair, compute the Pearson correlation for this realisation, and repeat the process 50 times. Doing so gives Pearson correlation values of $0.46\pm0.10$, $0.32\pm0.08$ and $0.36\pm0.08$, where the stated value and uncertainty are the mean and dispersion of the 50 realisations, for the diagnostic pairs PPO3N2-PPN2, K19N2O2-PPN2, and PP03N2-K19N2O2, respectively. This confirms that the correlation lengths derived from different diagnostics show reasonable consistency. The most likely explanation for the level of disagreement we do find is that the PP04 diagnostics both depend strongly on ionization parameter, which may vary systematically from galaxy to galaxy and across the face of each individual galaxy; by contrast, our default choice of K19N2O2 is quite insensitive to ionisation parameter. However, these ionisation parameter effects clearly have only minor effects on the statistical distribution of correlation lengths, as illustrated by \autoref{fig:corner}. Moreover, we have verified that, if we repeat the analysis presented in the remainder of this paper using correlation lengths derived from PPN2 or PPO3N2 rather than our default choice of K19N2O2, the change in results is in all cases within our estimated uncertainties.

\subsection{Correlation length as a function of galaxy physical properties}
\label{subsec:mass_sfr}

\begin{figure}
\includegraphics[width=1.0\linewidth]{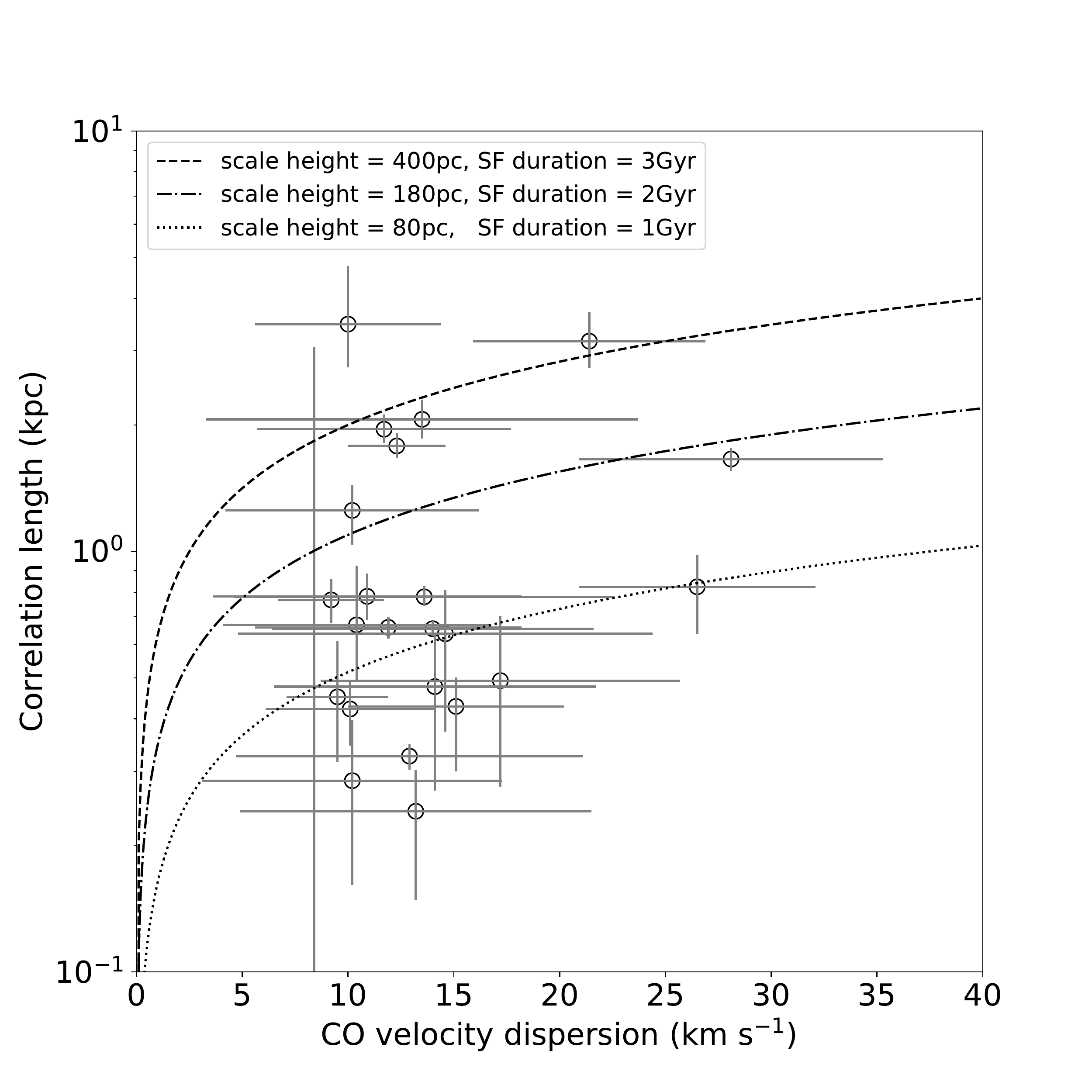}
\caption{Correlation length $l_{\rm corr}$ vs.~CO velocity dispersion $\sigma_{\rm g}$. The three different dashed curves show the values predicted by \autoref{equ:lcorr_2}, evaluated for the indicated $h$ and $t_*$. In the horizontal direction, points and error bars indicate the measured velocity dispersion and stated uncertainty from EDGE-CALIFA, while in the vertical direction the points and error bars indicate the 50th percentile value and 16th to 84th percentile range of the posterior PDF determined from our MCMC fit.}
\label{fig:vel_disp}
\end{figure}

Since the correlation length is an integral parameter of a galaxy, representing the level of diffusion in the ISM, the relation of the correlation length with other galaxy characteristic properties may provide some insights into galactic evolution. In the upper row of \autoref{fig:mass_sfr}, we compare the correlation length with stellar mass, SFR and $\mathrm{R}_{25}$. As with \autoref{fig:corner}, the quantity we plot is a heat map derived from bootstrap resampling over the posterior probability distributions of the variables plotted on the horizontal and vertical axes, and thus properly reflects the error distributions; we also show the typical sizes of the errors in the lower corner of each panel. It is clear from the figure that $l_\mathrm{corr}$ is positively correlated with stellar mass, SFR, and galactic radius. Quantitatively, the Pearson correlations between $l_{\rm corr}$ and stellar mass, SFR, and $\mathrm{R}_{\rm 25}$ (again derived from bootstrap resampling over the error distributions) are $0.30\pm 0.09$, $0.29\pm 0.06$, and $0.36\pm 0.07$, respectively. Broadly speaking, dwarf galaxies ($M_* < 10^{10}$ M$_\odot$, $\mbox{SFR} < 1$ M$_\odot$ yr$^{-1}$, $\mathrm{R}_{\rm 25} < 5$ kpc) have typical correlation lengths of $\sim 300$ pc, while spiral galaxies $(M_* > 10^{10}$ M$_\odot$, $\mbox{SFR}>1$ M$_\odot$ yr$^{-1}$, $\mathrm{R}_{\rm 25} > 5$ kpc) have typical correlation lengths of $\sim 1$ kpc.

Since galactic SFR, mass, and radius are all correlated with each other, it is interesting to attempt to disentangle which of them is the most closely related to $l_{\rm corr}$; the Pearson correlation is highest with $\mathrm{R}_{\rm 25}$, but by an amount that is comparable to the errors. To investigate this question further, we can ask whether there is any correlation between $l_{\rm corr}$ and stellar mass or star formation rate if we normalise by galaxy size. We answer this question in the panels in the lower row of \autoref{fig:mass_sfr}, where we show $l_{\rm corr}$ as a function of the stellar surface density $\Sigma_* \equiv M_*/\pi \mathrm{R}_{25}^2$ and the star formation surface density $\Sigma_{\rm SFR} \equiv \mathrm{SFR}/\pi \mathrm{R}_{25}^2$. These panels show no significant trends of $l_\mathrm{corr}$ with either stellar mass or star formation surface density, and the Pearson correlations are consistent with zero -- $0.09\pm 0.07$ for stellar surface density, $0.04\pm 0.07$ for gas surface density. Taken together, these results favour the hypothesis that the primary determinant of $l_{\rm corr}$ is simply the overall size of the galaxy. However, the physical mechanism that relates correlation length to physical scale of disc is not entirely clear. One possibility is that correlations are generated by large-scale organised structures in the disc such as spiral arms and bars, which might be larger in larger discs. We will explore the question of the physical origin of variations in the correlation length in more detail below.

\subsection{Correlation length and velocity dispersion}
\label{subsec:vel_disp}

One possible driver for the correlation between SFR and $l_{\rm corr}$ could be the increased velocity dispersion expected with higher SFRs \citep{Krumholz18}. The spectral resolution of the CALIFA sample is not high enough to test this proposition. Additionally the ionised gas velocity dispersions of the galaxies at lower star formation rates will be dominated by combined thermal and non-thermal velocities of $\sim 15$ km s$^{-1}$ that are internal to individual H~\textsc{ii} regions, rather than reflecting the velocity dispersion of the underlying ISM \citep[see Appendix B of][]{Krumholz18}. However, we can obtain more accurate velocity dispersions by cross-matching our sample with the EDGE-CALIFA survey \citep{Levy18}, which provides velocity dispersions based on CO measurements. We show the relation between $l_{\rm corr}$ and CO-derived velocity dispersion $\sigma_{\rm g}$ in \autoref{fig:vel_disp}. Here the horizontal errors bars show the reported uncertainties from EDGE-CALIFA, while the vertical error bars show the 16th to 84th percentile range of our posterior PDFs for $l_{\rm corr}$; points are placed at the 50th percentile.

Clearly there is no strong correlation evident in \autoref{fig:vel_disp}, but this is perhaps not surprising given the very small range in velocity dispersion covered by the sample. It is nonetheless interesting to compare the numerical results with general theoretical expectations. For a turbulent mixing process, the expected diffusion coefficient is $\kappa \sim l_{\rm turb} \sigma_{\rm g}/3$, where $l_{\rm turb}$ is the outer scale of the turbulence and $\sigma_{\rm g}$ is the gas velocity dispersion. The correlation length is therefore expected to be (KT18, their equation 91)
\begin{equation}
\label{equ:lcorr_2}
l_{\rm corr} = \sqrt{\frac{1}{3}h \sigma_{\mathrm{g}} t_*},
\end{equation}
where $h$ is the typical ISM scale height (which we take as an estimate of the outer scale of the turbulence) and $\sigma_{\mathrm{g}}$ represents the CO velocity dispersion. The three dashed curves in \autoref{fig:vel_disp} show the correlation length evaluated from \autoref{equ:lcorr_2} with three different combinations of $h$ and $t_*$. Scale heights are not directly measured for the EDGE-CALIFA galaxies, but the galaxies in the sample are relatively similar to the Milky Way, for which the scale height of the neutral ISM is $\approx 100 - 150$ pc, depending on the galactocentric radius \citep{Boulares90a, Wolfire03a}; we therefore plot a conservative range of a factor of $\sim 2-3$ about this value. Clearly the predicted values of $l_{\rm corr}$ are in agreement with the obseved ones only if the typical duration of star formation, $t_*$, is of order a few Gyr.

This timescale is obviously shorter than the actual ages of these galaxies ($\sim10$ Gyr). Since \autoref{equ:lcorr_2} comes from a fairly simple model, assuming  constant SFR, it is not surprising that the $t_*$ required to fit the data is not the actual galactic age. We can instead interpret $t_*$ as indicating something like the ``relaxation time'' over which the ISM settles to equilibrium. The estimated timescale is comparable to the gas depletion time of $\sim2$ Gyr \citep[e.g.,][]{Leroy13a}, and consistent with 1-4 Gyr time over which the radial metallicity gradient reaches equilibrium proposed by \citet{Sharda21}.

\begin{figure}
\includegraphics[width=1.0\linewidth]{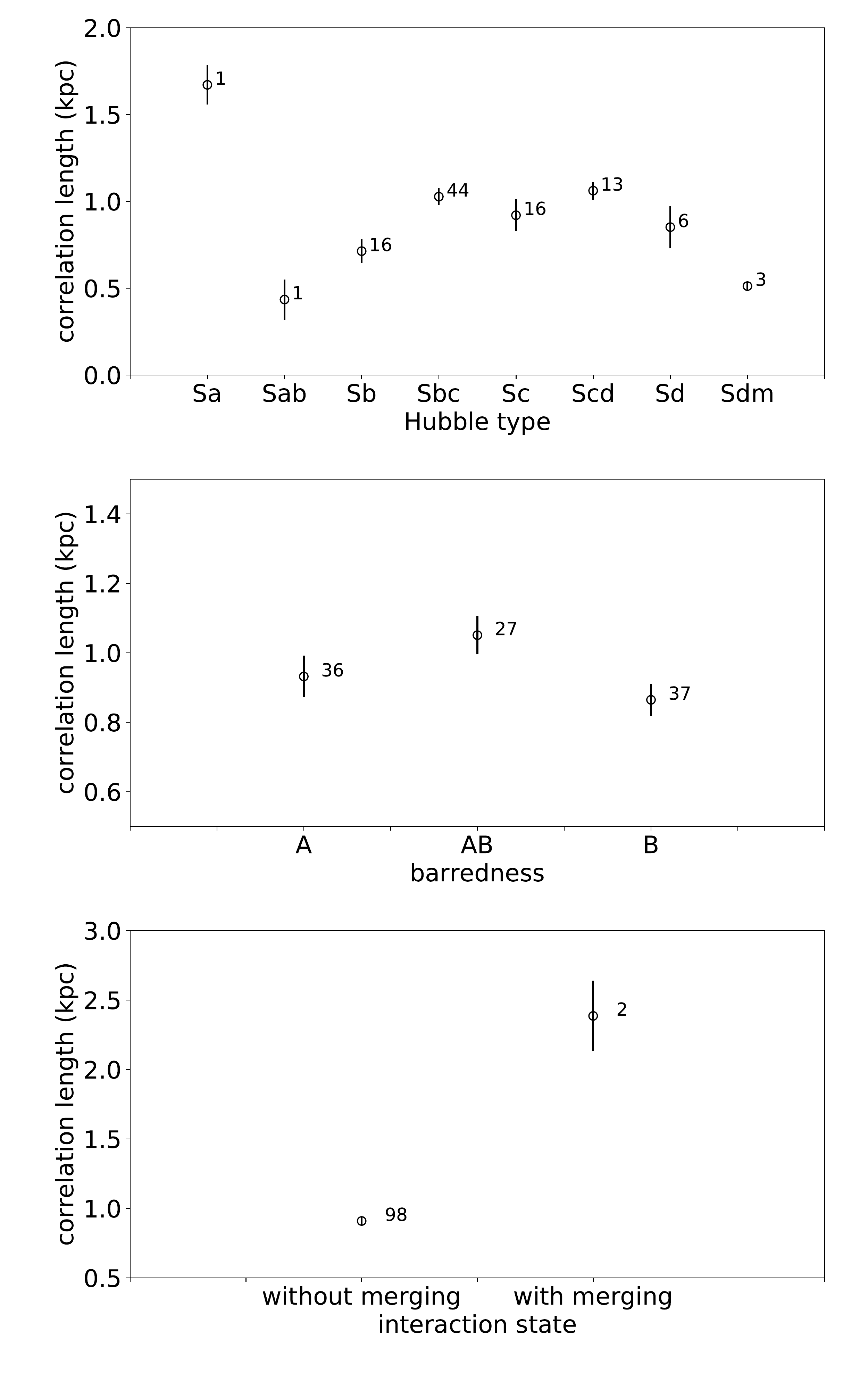}
\caption{Correlation lengths of galaxies grouped by Hubble type (top), barredness (middle) and interaction state (bottom). The sample size is labeled for each circle. For each point shown, we compute the mean and standard deviation using 50-realisation bootstrapping. We draw a random $l_{\rm corr}$ for each galaxy from the galaxy's posterior PDF. For those galaxies in one group, we compute one mean value of $l_{\rm corr}$. We repeat this procedure for 50 times and get 50 mean values. The mean and standard deviation of these 50 mean values are our plotted circle and error bar in one group. In the middle panel, ``A'' indicates no bar, ``B'' is barred, and ``AB'' indicates a weak or ambiguous bar.}
\label{fig:morph}
\end{figure}

\begin{figure}
\includegraphics[width=1.0\linewidth]{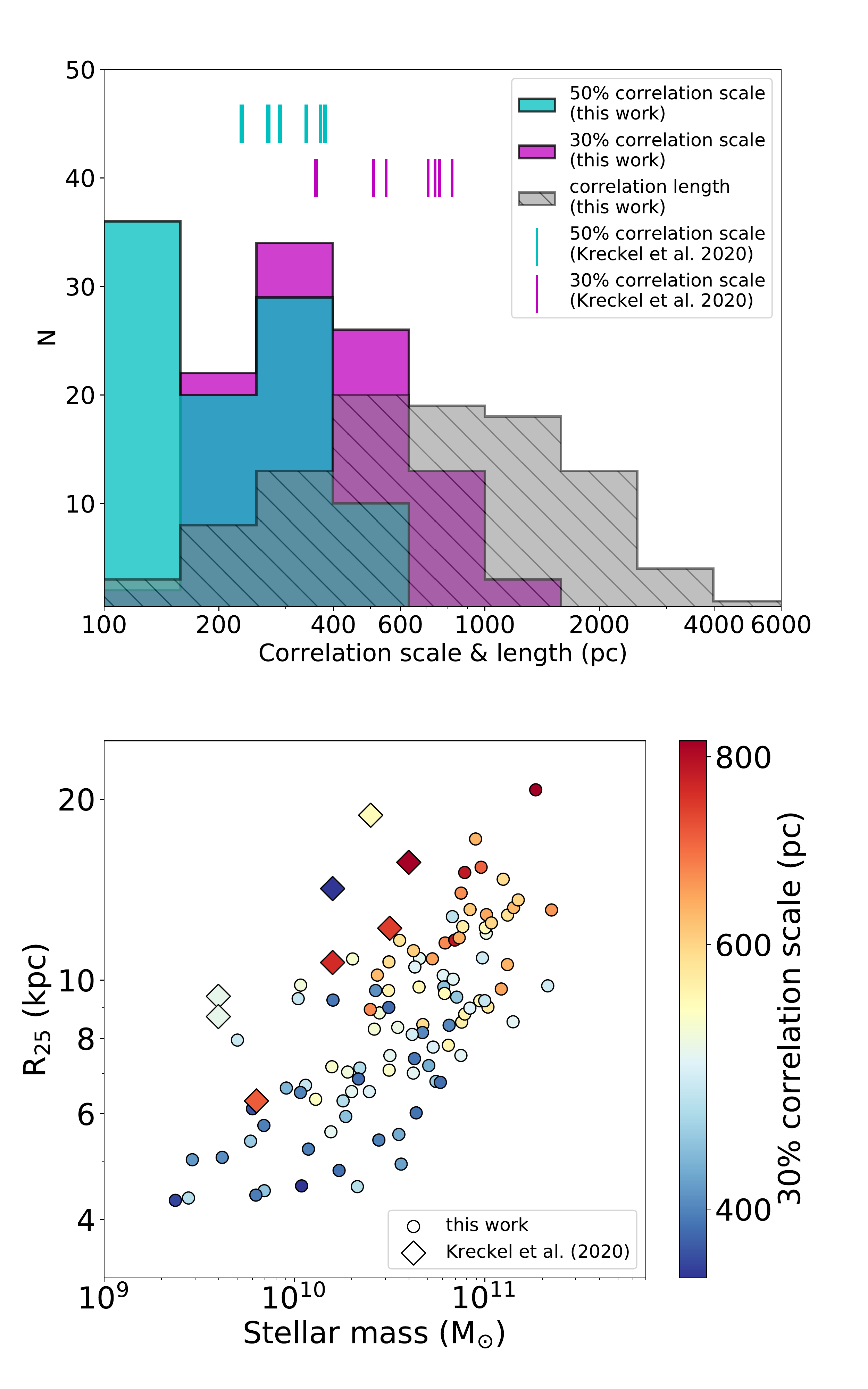}
\caption{Upper: distribution of 50\% and 30\% correlation scales of our sample, defined as the length scales $r$ for which $\xi(r) = 0.5$ and $0.3$, respectively (cyan and magenta histograms). We show the corresponding measurements for the 8 galaxies analysed by \citet{Kreckel20} as coloured vertical lines, with the upper (cyan) and lower (magenta) rows indicating the 50\% and 30\% correlation scales, respectively. Lower: the distribution of our sample (circles) and \citeauthor{Kreckel20}'s sample  (diamonds), with colour indicating the 30\% correlation scale.}
\label{fig:corr_scale}
\end{figure}

\subsection{Correlation length and morphological parameters}
\label{subsec:morph}

Several studies \citep[e.g.][]{Sanchez-Menguiano16, Ho18, Spitoni19} propose that large-scale structures within galaxy discs (bars, spiral arms) may play an important role in ISM mixing. Our finding that larger galaxies have larger correlation scales is at least qualitatively consistent with this idea. We therefore investigate whether correlation lengths vary with galaxy morphological characteristics in our sample, using the morphological assignments provided by \citet{Walcher14}. In \autoref{fig:morph}, we show the correlation lengths of galaxies grouped by Hubble type, barredness, and interaction state. We find no significant trend of $l_{\rm corr}$ from Sa to Sd galaxies in our sample, though it is possible that this mostly reflects the small size our sample at the extreme ends of the Hubble type range. It would be interesting to explore this further with larger samples \citep{Colombo18}. Similarly, although we find that merging galaxies have longer correlation lengths, the sample size is only two galaxies and we therefore do not regard this result as particularly convincing. Galaxies with different barredness show almost identical (within $2\sigma$) $l_{\rm corr}$. We therefore find no strong trends relating galactic correlation lengths to large-scale morphological features. Further comparison between our results and models and simulations of bar-driven or spiral-arm-driven mixing \citep[e.g.][]{DiMatteo13, Grand16, Sanchez-Menguiano16} are beyond the scope of our work.

\section{Comparison with previous work }
\label{sec:comparison}

To date the only prior published measurements of metallicity correlations in galaxies are from \citet{Kreckel20}, who report the two-point correlations of eight nearby galaxies observed by the PHANGS survey. \citeauthor{Kreckel20} report the lags for which the correlation falls to 30\% and 50\%, which they find to be typically around 500 pc. To compare to this work, we compute the same lags for our sample, using a simple linear interpolation of our binned two-point correlation functions (\autoref{fig:TPCF}). We compare the distribution of 30\% and 50\% lags for our sample to compare to those reported by \citeauthor{Kreckel20} in \autoref{fig:corr_scale}. While we cannot perform a detailed statistical comparison of the distributions given the small size of the PHANGS sample, it is clear that the distributions are qualitatively similar - nearby galaxies typically have $\sim200$-pc 30\% correlation scale and $\sim300$-pc 50\% correlation scale. Similarly, both of our results show that the distribution of 50\% correlation lengths is more tightly clustered than that of 30\% correlation lengths. Our correlation scales are slightly smaller on average than \citeauthor{Kreckel20}'s, and this difference may be due to the fact that \citeauthor{Kreckel20}'s galaxies have larger radii (ranging from $\sim6$ kpc to $\sim19$ kpc, typically $\sim12$ kpc), while our sample contains smaller galaxies (see the middle panel of \autoref{fig:basic}), which have smaller $l_{\rm corr}$ in general (see \autoref{subsec:mass_sfr} and \autoref{fig:mass_sfr}). We can see this effect clearly in the lower panel of \autoref{fig:corr_scale}, where we compare the two samples and the 30\% correlation lengths derived from them in the stellar mass - radius plane. Consistent with our analysis in \autoref{subsec:mass_sfr}, \autoref{fig:corr_scale} shows that larger-size galaxies produce larger correlation scales. It also shows that, at a given point in mass-radius space, our sample has 30\% correlation scales that are qualitatively similar to \citeauthor{Kreckel20}'s; thus the slightly larger average value of the 30\% correlation scale in \citeauthor{Kreckel20}'s sample is most likely just a reflection of the differing galaxy selections in PHANGS and CALIFA.

While the data are similar, however, we note that simply defining a correlation scale (e.g. 30\% or 50\%) from the correlation function is not necessarily the most direct way to investigate the physical processes of interest. In the KT18 model for example, these scales are influenced by both $\sigma_0$ (which combines the intrinsic width of injection events and the effects of beam smearing) and $\kappa t_*$ (which captures the effects of diffusion); in addition to these physical factors, the measured two-point correlation is affected by the level of uncertainty in the metallicity measurements (captured by the parameter $f$ in our model fits -- see \autoref{subsec:MCMC}). Consequently, one can reproduce a certain 30\% correlation scale by either modifying $\sigma_0$ (which mostly stretches the curve describing $\xi(r)$ in $r$), by modifying $\kappa t_*$ (which mostly raises or lowers $\xi(r)$ at fixed $r$, see Figure 1 in KT18), or by altering $f$ (which purely increases or decreases $\xi(r)$ for all $r$ larger than a single observed pixel). A single measurement of the length scale at which the correlation is 30\% or 50\% is incapable of disentangling these possibilities. Our parametric fit, though necessarily model-dependent, avoids this problem because it makes use of the full shape of the $\xi(r)$ curve.

Our measured correlation lengths $l_{\rm corr}$ are typically somewhat larger than even the 30\% correlation scale: $\sim 1$ kpc versus $\sim 0.4$ kpc (compare the histograms in the upper panel of \autoref{fig:corr_scale}). This is partly due to the fact that $l_{\rm corr}$ is not exactly the same as the 30\% or 50\% correlation length, but a larger factor is that our parametric fits account for the effects of uncertainty in metallicity diagnostics. As discussed in \autoref{subsec:MCMC} and \aref{app:kt18_generalisation}, the effect of errors in the metallicity diagnostic is to reduce the measured correlation function for all lags larger than a single pixel, which in turn reduces the inferred length scale at which the correlation falls to some specified value such as 30\%. Thus a correlation scale taken directly from the measured two-point correlation, rather than extracted from a model fit that accounts for errors in pixel metallicities, gives a lower limit on the true correlation length. This is why, although our measured distributions of the 30\% and 50\% correlation scale are qualitatively similar to, or even slightly smaller than, those reported by \citet{Kreckel20}, our inferred correlation lengths are a factor of $\sim 2.5$ larger.

\section{Conclusions}
\label{sec:conclusions}

We make use of CALIFA DR3 Pipe3D emission line maps to analyse the two-point correlation functions of the metallicity distribution in 100 nearby galaxies. After removing the radial metallicity gradients, we detect significant correlations (i.e., correlations well in excess of what would be expected due to beam-smearing alone) in essentially all the galaxies in the sample. Our results establish the principle that the 2D metallicity distributions in galactic discs carry significant statistical information. We further show that the metallicity correlations we measure are well-fit by a simple injection-diffusion model as proposed by \citet{KT18}. The most important parameter in this model is the correlation length ($l_{\rm corr}$), which describes the typical length scale over which diffusion of metals induces correlations in the metallicity distribution across the face of a galaxy disc. We show that this quantity is strongly constrained by the observed metallicity correlation functions, and has a mean value of $\sim1$ kpc in our sample.

We also investigate the relationships between $l_{\rm corr}$ and other galaxy parameters. We find that $l_{\rm corr}$ is positively correled with both stellar mass and star formation rate, though it is unclear which of these quantities (if either) is the true driver of the correlation, since stellar mass and star formation rate are also correlated with each other. We also find no significant trends between $l_{\rm corr}$ and morphological parameters such as Hubble type, bar strength, or galaxy interaction, though in some cases these conclusions are weak due to limited sample sizes. Finally, we search for a correlation between $l_{\rm corr}$ and gas velocity dispersion predicted by the \citet{KT18} model; for this purpose we make use of the velocity dispersions measured in a subset of our sample by the EDGE-CALIFA survey \citep{Levy18}. However, because the sample covers only a narrow range of velocity dispersion ($\approx 10-25$ km s$^{-1}$), out statistical power is limited and we find no strong evidence for correlation.

In future work we intend to extend this technique to additional galaxy samples. In one direction, we can use higher-resolution surveys such as the Muse Atlas of Discs \citep[MAD;][]{Erroz-Ferrer19}, which, while limited in sample size, will allow us to probe the two-point correlation on significantly finer scales. This will provide a stronger test of whether the simple \citet{KT18} model provides an adequate description of the metallicity structure, or whether a more complex model is required, for example one that explicitly accounts for large-scale galactic structures such as spiral arms. In another direction, our detection of correlation lengths as large as $\sim 1$ kpc suggests that it would be worthwhile to repeat this analysis using other galaxy surveys that offer somewhat worse spatial resolution than CALIFA, but also larger sample sizes. This might sharpen our view of the relationship between correlation length and galaxy mass we have discovered in this work, and might reveal additional trends with galaxy properties that are too small to be seen in our current sample.

\section*{Acknowledgements}

We thank the anonymous reviewer for carefully reading the
manuscript and providing constructive comments and suggestions. We acknowledge the observations collected at the Centro Astron\'omico Hispano-Alem\'an (CAHA) at Calar Alto, operated jointly by Junta de Andalucía and Consejo Superior de Investigaciones Científicas (IAA-CSIC). MRK acknowledges support from the Australian Research Council through awards FT180100375 and DP190101258. EW \& JTM acknowledge support by the Australian Research Council Centre of Excellence for All Sky Astrophysics in 3 Dimensions (ASTRO 3D), through project number CE170100013. SFS is grateful for the support of a CONACYT grant CB-285080 and FC-2016-01-1916, and funding from the PAPIIT-DGAPA-IN100519 (UNAM) project. L.G. was funded by the European Union's Horizon 2020 research and innovation programme under the Marie Sk\l{}odowska-Curie grant agreement No. 839090. This work has been partially supported by the Spanish grant PGC2018-095317-B-C21 within the European Funds for Regional Development (FEDER). ZL thanks Yifei Jin for fruitful discussion and generous support. ZL also thanks Yankun Di for the detailed discussion of auto-correlation function of finite and non-periodic space and time sequences.

\section*{Data Availability Statement}

The data and the source code underlying this article are available at \url{https://doi.org/10.5281/zenodo.4650718}.

%%%%%%%%%%%%%%%%%%%%%%%%%%%%%%%%%%%%%%%%%%%%%%%%%%



\bibliographystyle{mnras}
\bibliography{ref}

%%%%%%%%%%%%%%%%%%%%%%%%%%%%%%%%%%%%%%%%%%%%%%%%%%

%%%%%%%%%%%%%%%%% APPENDICES %%%%%%%%%%%%%%%%%%%%%

\appendix

\section{Dependence of the results on AGN pixel masking}
\label{app:agn}

\begin{figure}
\includegraphics[width=1.0\linewidth]{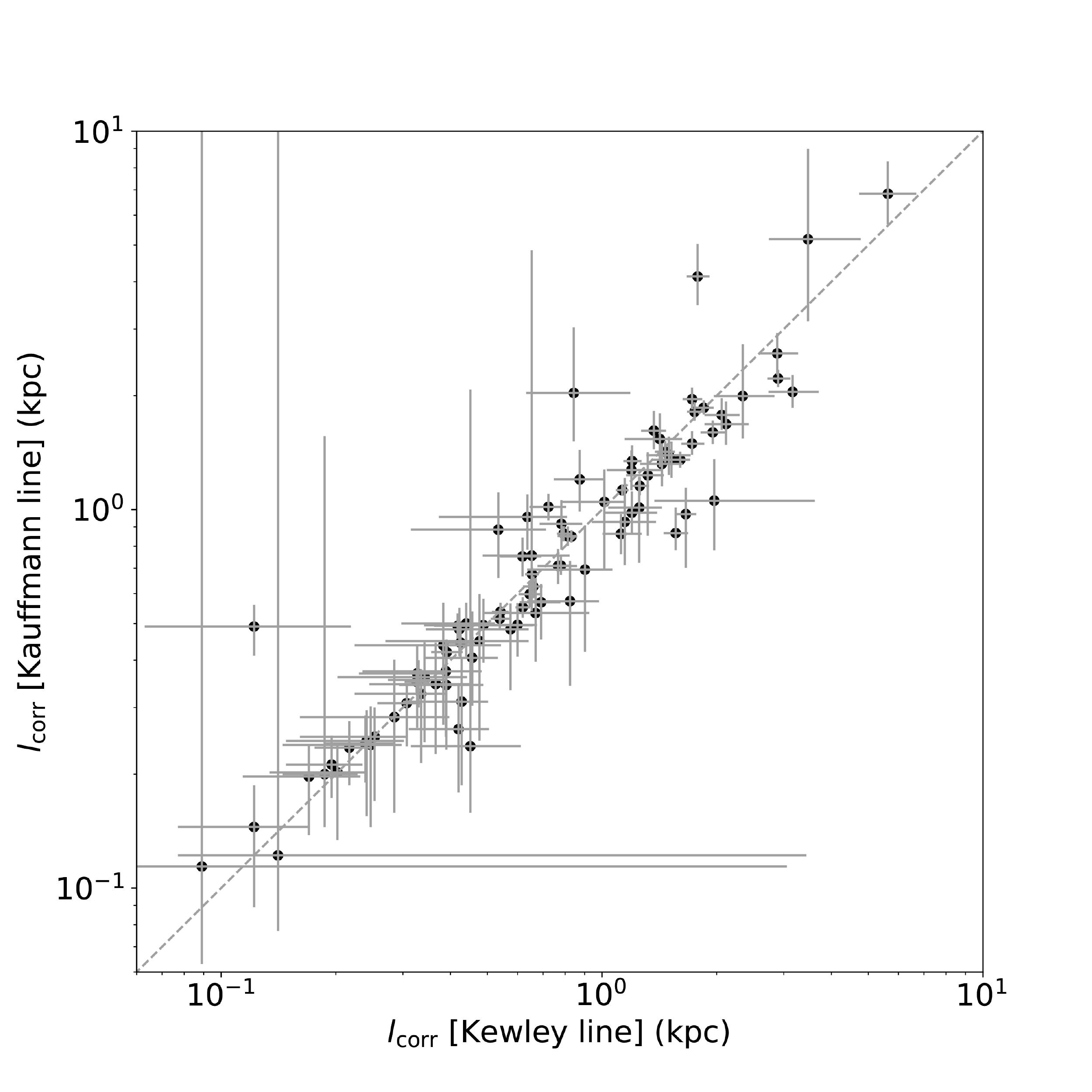}
\caption{Comparison of correlation length derived using Kewley line versus Kauffman line. Points show the 50th percentile values for each galaxy, and error bars show the 16th to 84th percentile range. The Pearson correlation between the two sets, derived via bootstrap resampling of the posterior PDFs (see \autoref{subsec:tests}) is $0.58\pm0.17$, which shows significant consistency.}
\label{fig:agn}
\end{figure}

In the analysis presented in the main paper, we mask pixels for possible AGN contamination using the \citeauthor{Kewley01} criterion, which attempts to retain pixels that contain no more than a $\approx 20\%$ contribution from shocks or AGN. An alternative approach would be to mask based on the stricter \citet{Kauffmann03}, which attempts to screen out a lower level of AGN contribution. We choose the more relaxed criterion because the [N~\textsc{ii}]/[O~\textsc{ii}]-based K19N2O2 metallicity diagnostic that we use is not sensitive to $\lesssim 20\%$ AGN or shock contamination \citep{KE08}. (However, this is not true for the PP04 diagnostics to which we compare in \autoref{subsec:tests}, which are very sensitive to even a 20\% contamination from AGN.) To ensure that our results are not overly sensitive to the AGN masking criterion, we carry out an additional experiment by analysing our sample using the \citeauthor{Kauffmann03} rather than the \citeauthor{Kewley01} criterion to mask AGN-contaminated pixels. \autoref{fig:agn} shows the comparison of $l_{\rm corr}$ derived using the two different AGN criteria for the full 100-galaxy sample. It illustrates that switching the Kewley line to the Kauffmann line does not significantly influence the $l_{\rm corr}$ for the K19N2O2 diagnostic. The main effect of using the stricter Kauffmann line is to reduce the number of pixels available for analysis by enlarging the central ``halo'' of masked spaxels near galactic centres. This slightly changes the two-point correlation function, and leads to larger error bars, since there are fewer pixels. However, \autoref{fig:agn} illustrates that these effects are small. We therefore adopt Kewley line in our analysis.

\section{Comparison of results derived from CALIFA DR3 imaging with previous results}
\label{app:decom}

\begin{figure*}
\includegraphics[width=1.0\linewidth]{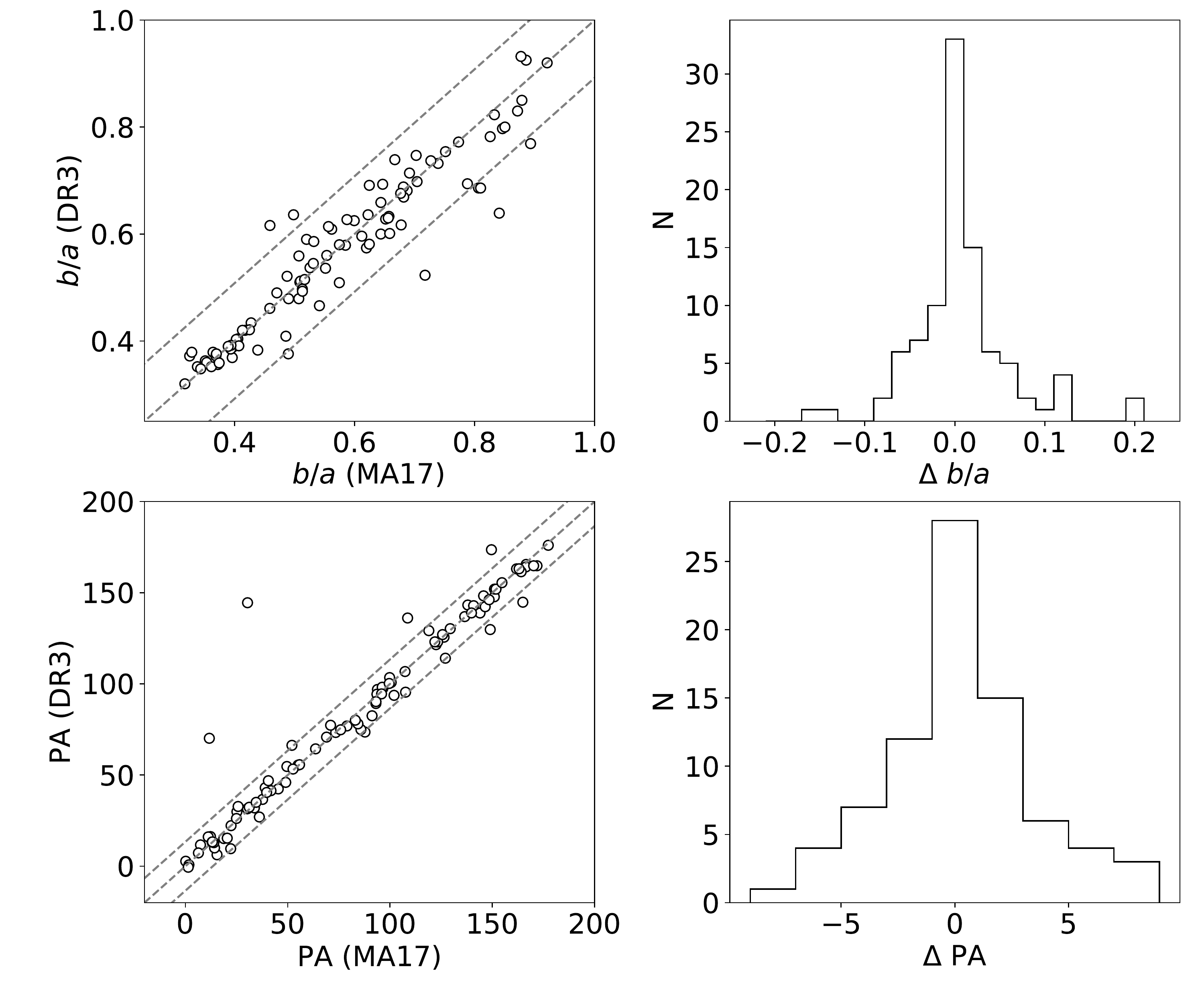}
\caption{Comparison of axis ratio (top) and position angle (bottom) derived from CALIFA DR3 image pipeline versus \citet{MA17}. The two right panels show the distribution of the differences $\Delta$ in the axis ratio and position angle between the two samples. The three dashed lines in each panel on the left show 1-1 line and the 2$\sigma$ range around it.}
\label{fig:decom}
\end{figure*}

\begin{figure}
\includegraphics[width=1.0\linewidth]{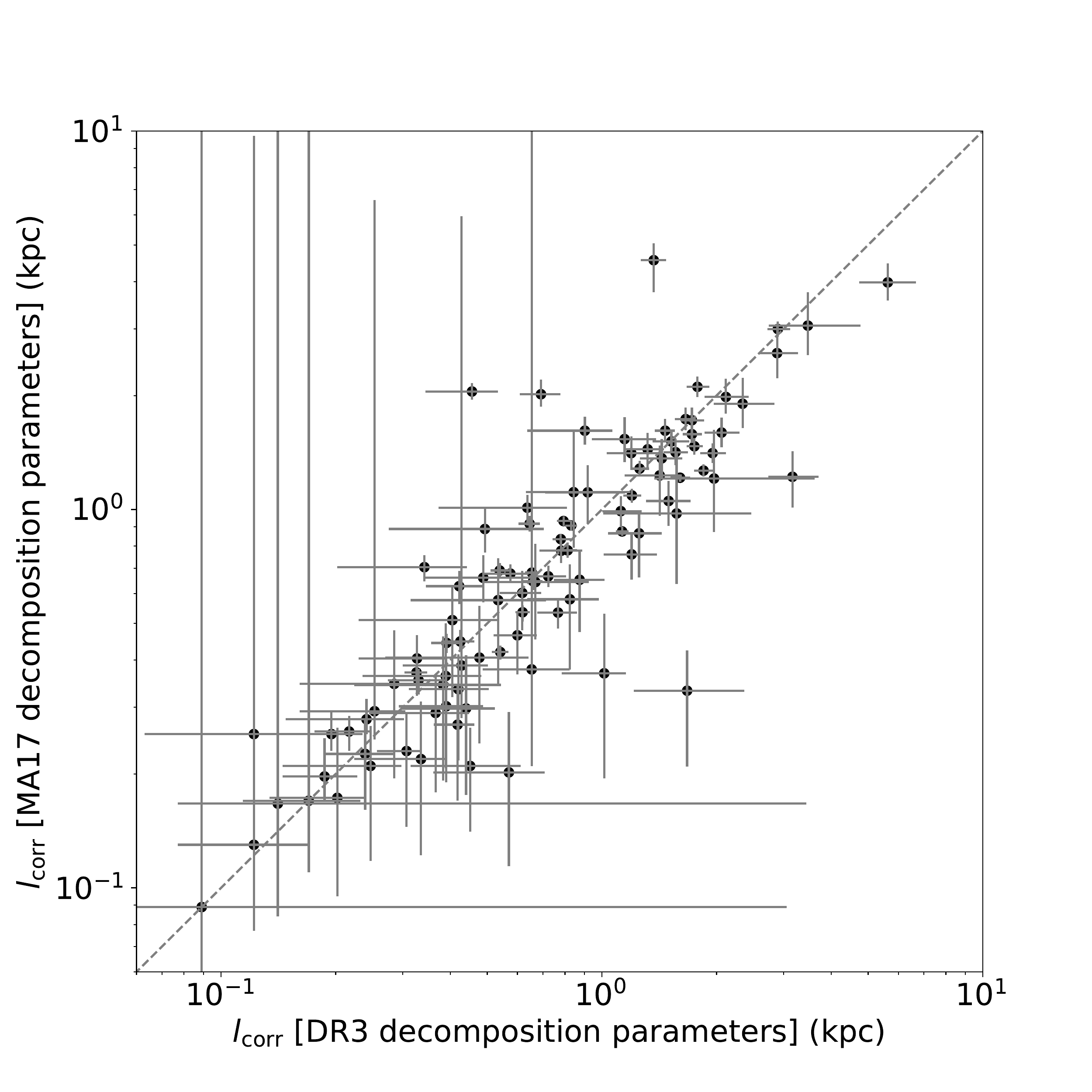}
\caption{Comparison of correlation length derived using DR3 image decomposition parameters versus parameters taken from the published image decomposition of MA17. Points show the 50th percentile values for each galaxy, and error bars show the 16th to 84th percentile range. The Pearson correlation between the two sets, derived via bootstrap resampling of the posterior PDFs (see \autoref{subsec:tests}) is $0.51\pm0.14$, which shows significant consistency.}
\label{fig:_decom}
\end{figure}

Our analysis makes use of position angles (PA), ellipticities ($\epsilon$) and inclination angles ($i$) taken from the most recent results from CALIFA DR3, which decomposes galaxy images using an updated image pipeline that combines the SDSS, PanSTARRS, and DES surveys. The values we use are listed in our \autoref{tbl:basic}.
%but because the underlying data have not yet been described in the literature, it is important to compare them to earlier published results.
\citet[][hereafter MA17]{MA17} report a two-dimensional multi-component photometric decomposition of 404 galaxies drawn from CALIFA DR3. We compare the updated values for the axis ratio ($b/a$) and position angle to the MA17 results in \autoref{fig:decom}. We find that for 95\% of galaxies in our sample, the adopted value for the axis ratio differs from \citeauthor{MA17}'s by $<0.1$, and the adopted value for the PA differs by $<13^{\circ}$. We have also re-run the full analysis pipeline using values from MA17. \autoref{fig:_decom} shows the comparison of $l_{\rm corr}$ derived using the two different sets of parameters. From the figure, we see that for the great majority of our sample, the derived $l_{\rm corr}$ values agree within the error bars, and the estimated Pearson correlation between the two sets of estimates, derived from bootstrap resampling following the procedure outlined in \autoref{subsec:tests}, is $0.51\pm 0.14$.

\section{Dependence of results on the size of annular bins}
\label{app:adp}

\begin{figure}
\includegraphics[width=1.0\linewidth]{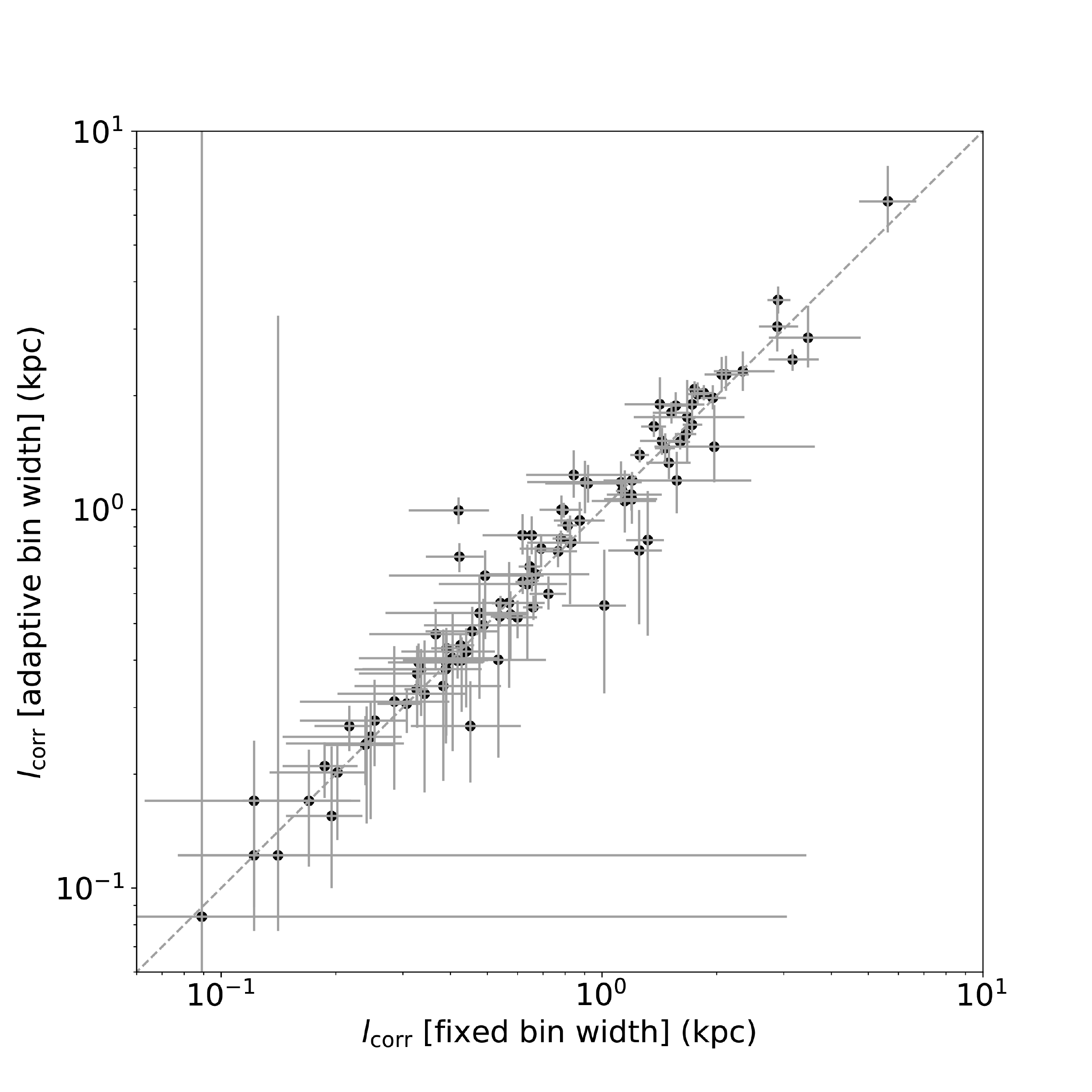}
\caption{Comparison of correlation length derived using fixed versus adpative bin widths. Points show the 50th percentile values for each galaxy, and error bars show the 16th to 84th percentile range. The Pearson correlation between the two sets, derived via bootstrap resampling of the posterior PDFs (see \autoref{subsec:tests}) is $0.79\pm0.07$, which shows significant consistency.}
\label{fig:adp}
\end{figure}

In the analysis presented in the main paper, we adopt a fixed bin width (0.2 kpc) for the purpose of computing the mean metallicity in each annulus, which we subtract in order to remove the radial metallicity gradient. To ensure that this choice does not substantially affect our results, We repeat our analysis using adaptive bin widths as computed by the numpy histogram function set to use optimal binning. \autoref{fig:adp} shows the comparison of $l_\mathrm{corr}$ derived using the two different binning algorithms for the full 100-galaxy sample. It shows that changing the binning method does not significantly influence the $l_\mathrm{corr}$. Thus, we adopt a 0.2 kpc fixed bin width in our analysis.

\section{Generalisation of the KT18 model to include observational effects}
\label{app:kt18_generalisation}

In this appendix we generalise the KT18 model for the correlation function of the metallicity in a galactic disc to account for the effects of noise and finite observational resolution. KT18 idealise the metallicity distribution in a galaxy as arising from the competition between metal ejection events, which are uniformly-distributed in space and time and add metals in a small area, and a diffusion process, which smooths out the metal field. The problem is described by five dimensional parameters: the rate of injection events per unit time per unit area, $\Gamma$, the dispersion of the spatial distribution (assumed to be Gaussian) of metals added in an injection event $\sigma_{\rm inj}$, the mean mass of metal injected per event $\langle m_X\rangle$, the diffusion coefficient $\kappa$, and the time $t_*$ since the start of injection. They show that the two-point correlation function of the resulting metal field is given by
\begin{equation}
\xi(r) = \frac{2}{\ln\left(1 + \tau_f/\tau_0\right)}
\int_0^\infty e^{-2 \tau_0 k^2} \left(1 - e^{-2\tau_f k^2}\right) \frac{J_0(kx)}{k} \, dk,
\end{equation}
where $J_n(x)$ is the modified Bessel function of the first kind of order $n$, and we have defined the dimensionless length and time scales
\begin{eqnarray}
& x = r \left(\frac{\Gamma}{\kappa}\right)^{1/4} \\
& \tau_f = t_*\sqrt{\Gamma\kappa} \\
& \tau_0 = \frac{\sigma_{\rm inj}^2}{2} \sqrt{\frac{\Gamma}{\kappa}}.
\end{eqnarray}

Now suppose that we observe this metal field with a finite spatial resolution, which we will approximate as being equivalent to convolving the metal field with a Gaussian beam of dispersion $\sigma_{\rm beam}$. To determine the correlation of the beam-convolved metal field, we follow KT18 by working in dimensionless variables where distances are measured in units of $(\kappa/\Gamma)^{1/4}$, times in units of $(\Gamma\kappa)^{-1/2}$, and metal masses are measured in units of $\langle m_X\rangle$. In this unit system, the power spectral density of the metallicity distribution is (KT18, equation 63):
\begin{eqnarray}
\lefteqn{
\Psi(k) = \frac{1+\sigma_w^2}{8\pi^2} e^{-2\tau_0 k^2}
\left(\frac{1-e^{-\tau_f k^2}}{k^2}\right)
\left\{1 + e^{-\tau_f k^2} 
+ {}
\phantom{\left[\frac{1 - \left(1+\tau_f k^2\right) e^{-\tau_f k^2}}{\tau_f k^2}\right]\hspace{-3in}}
\right.
}
\nonumber \\
& &
\qquad\qquad
\left.
\frac{8\pi}{1+\sigma_w^2} \left(\frac{1 - e^{-\tau_f k^2}}{ k^2}\right)
\left[\frac{J_1(k R')}{k}\right]^2
\right\}.
\label{equ:psd_full}
\end{eqnarray}
Here $\sigma_w$ is the dispersion in the total mass of metals injected by the ejection events (i.e., if all injection events add an equal mass of metals, $\sigma_w = 0$), the value of which will prove not to matter as long as it is finite, and $R'$ is a dummy variable describing the size of the region over which the metals are injected, which we will take to infinity -- see KT18 for details. The two-point correlation function is related to the power spectral density by (equations 44 and 48 of KT18)
\begin{equation}
\label{equ:2point}
\xi(x) = \frac{1}{\sigma^2}\left[2\pi \int_0^\infty \Psi(k) J_0(k x) k \, dk - \tau_f^2\right],
\end{equation}
where $\sigma^2$ is the total variance of the (dimensionless) metal field, which we denote $S_X$.

We can compute the power spectral density of the beam-convolved metal field by noting that, in Fourier space, convolution with such a Gaussian is equivalent to multiplication by the kernel function
\begin{equation}
K(k) = e^{-k^2 s_{\rm beam}^2/2},
\end{equation}
where $s_{\rm beam} = \sigma_{\rm beam} (\Gamma/\kappa)^{1/4}$,
so the power spectral density of the beam-convolved metal field is
\begin{eqnarray}
\lefteqn{
\Psi_{\rm conv}(k) = \frac{1+\sigma_w^2}{8\pi^2} e^{-k^2/2s_{\rm beam}^2} e^{-2\tau_0 k^2}
\left(\frac{1-e^{-\tau_f k^2}}{k^2}\right) \cdot {}
}
\nonumber \\
& &
\!\!\!\!\!\!\!\!\!
\left\{1 + e^{-\tau_f k^2} 
+ 
\frac{8\pi}{1+\sigma_w^2} \left(\frac{1 - e^{-\tau_f k^2}}{ k^2}\right)
\left[\frac{J_1(k R')}{k}\right]^2
\right\}.
\label{equ:psd_obs}
\end{eqnarray}
The two-point correlation function of the convolved field is then given by \autoref{equ:2point} with $\sigma^2$ and $\Psi(k)$ replaced by the variance $\sigma_{\rm conv}^2$ and power spectral density $\Psi_{\rm conv}(k)$ of the beam-convolved field.

To compute the variance of the convolved metal field, we note that it is related to the mean $\langle S_X\rangle$ and mean square $\langle S_X^2\rangle$ of the metal field by
\begin{equation}
\sigma^2 = \langle S_X^2\rangle - \langle S_x\rangle^2,
\end{equation}
where the angle brackets indicate areal averages. KT18 show that the mean value of the dimensionless metal field $\langle S_X\rangle = \tau_f$, and this is unaffected by beam convolution, since convolution can only change the apparent distribution of metals, not the total quantity. We can obtain the mean square value of the metal field using the Wiener-Khinchin Theorem
\begin{equation}
\langle S_X^2\rangle = 2 \pi \int \Psi(k) k \, dk.
\end{equation}
Applying this to the un-convolved distribution, and taking the limit $R'\rightarrow \infty$ so that the region over which metal injection occurs is much larger than the scales over which we are computing the correlation, gives
\begin{equation}
\sigma^2 = \frac{1+\sigma_w^2}{8\pi} \ln\left(1 + \frac{\tau_f}{\tau_0}\right),
\end{equation}
which matches equation 38 of KT18.\footnote{In evaluating the terms in the integral proportional to $J_1(k R')$, we note that, in the limit $R'\rightarrow\infty$, this term becomes zero everywhere except in a small neighbourhood around $k=0$. We therefore replace all the terms that multiply $J_1$ by their leading-order Taylor expansions about $k=0$, which renders the integral analytically soluble. See KT18 for further discussion.} Applying the same treatment to the convolved metal field, the variance is
\begin{equation}
\sigma_{\rm conv}^2 = \frac{1+\sigma_w^2}{8\pi} \ln\left(1 + \frac{4 \tau_f}{s_{\rm beam}^2 + 4 \tau_0}\right).
\end{equation}
This clearly reduces to the un-convolved case in the limit $\sigma_{\rm beam}\rightarrow 0$, as it should.

Now computing the two-point correlation function of the convolved field using \autoref{equ:2point}, we have
\begin{eqnarray}
\lefteqn{\xi_{\rm conv}(r) = \frac{2}{\ln\left(1 + \frac{4\tau_f}{s_{\rm beam}^2+4\tau_0}\right)}
}
\nonumber \\
& &
\int_0^\infty e^{-2 (\tau_0 + s_{\rm beam}^2/4) k^2} \left(1 - e^{-2\tau_f k^2}\right) \frac{J_0(kx)}{k} \, dk.
\label{equ:corr_conv}
\end{eqnarray}
Comparing to Equation 67 of KT18, it is clear that this is equivalent to just taking the original KT18 result and replacing $\tau_0$ by $\tau_0 + s_{\rm beam}^2/4$, or, equivalently, replacing the squared injection width $\sigma_{\rm inj}^2$ with $\sigma_{\rm inj}^2 + \sigma_{\rm beam}^2/2$.

The expression given by \autoref{equ:corr_conv} incorporates the effects of finite spatial resolution, but not yet the effects of noise. We idealise the noise as causing a Gaussian random error in the observed metal field, such that, if the true (beam-convolved) metal field at a given position is $S_X$, then the observed metal field is described by a random variable centred on $S_X$ with some dispersion $s_{\rm err}$, which for simplicity we will take to be independent of position. We further assume that errors are spatially uncorrelated, i.e., the error in recovered metallicity at one position in the galaxy is uncorrelated with the error at any other position. The correlation function is (KT18, equation 44)
\begin{equation}
    \xi(r) = \frac{\left\langle S_X(\mathbf{x}+\mathbf{x}') S_X(\mathbf{x}')\right\rangle - \left\langle S_X\right\rangle^2}{\sigma^2},
\end{equation}
where $S_X$ and $\sigma$ here can be either the un-convolved or the convolved values, and the quantities in angle brackets are averaged over the dummy position variable $\mathbf{x}'$. Under our assumption of uniform, spatially uncorrelated noise, both of the terms in the numerator are left unchanged by the noise, since the noise terms will cancel to zero when averaged over space. On the other hand, the contribution of noise to the denominator $\sigma^2$ will not cancel, and will instead add in quadrature, i.e., if the true variance of the metal field is $\sigma^2$, and we measure with some Gaussian error $s_{\rm err}$, the observed variance will be $\sigma_{\rm obs}^2 = \sigma^2 + s_{\rm err}^2$. For convenience, we define $f = \sigma_{\rm obs}^2 / \sigma^2$, i.e., $f$ describes the factor by which noise increases the observed variance relative to the true value. With this definition, the effect of noise is simply to reduce the correlation by a factor $f$.
However, note that this reduction cannot hold for all lags as $r \to 0$, since then we would not have $\xi(0) = 1$; the problem is that our assumption of spatially-uncorrelated noise cannot to hold for arbitrarily small spatial scales, since the noise must be correlated at $r=0$ -- points always correlate perfectly with themselves! For a pixelised observation, the equivalent statement is that errors in the metallicities inferred for different pixels can be uncorrelated, but pixels always correlate perfectly with themselves. Thus the correlation function is reduced by a factor $f$ for all separations $>1$ pixel. With this assumption, and substituting back in our dimensional variables for convenience, we arrive at our final expression for the predicted correlation function:
\begin{eqnarray}
\lefteqn{\xi_{\rm model}(r) = \frac{2}{\ln\left(1 + \frac{2\kappa t_*}{\sigma_0^2/2}\right)
} \left[\frac{\Theta(r-\ell_{\rm pix})}{f} + \Theta(\ell_{\rm pix}-r)\right]
}
\nonumber \\
& &
\int_0^\infty e^{-\sigma_0^2 a^2/2} \left(1 - e^{-2 \kappa t_* a^2}\right) \frac{J_0(ar)}{a} \, da,
\label{equ:corr_conv_noise}
\end{eqnarray}
where $\sigma_0^2 = \sigma_{\rm beam}^2 + 2\sigma_{\rm inj}^2$, $a = k (\Gamma/\kappa)^{1/4}$, $\ell_{\rm pix}$ is the size of a pixel, and $\Theta(x)$ is the Heaviside step function, which is unity for $x>0$ and zero for $x<0$.

% Don't change these lines
\bsp	% typesetting comment
\label{lastpage}
\end{document}